\documentclass[a4paper,11pt]{article}
\pdfoutput=1 \usepackage{jheppub_2} 
\usepackage[citecolor=blue]{hyperref}
\usepackage{graphicx, multirow,soul,url,amsmath,amsfonts,amssymb,mathrsfs,amsfonts}
\usepackage[all]{hypcap}
\usepackage[T1]{fontenc} 
\usepackage[applemac]{inputenc}
\usepackage{url}
\usepackage{minibox}
\usepackage{bbold}
\usepackage{color}
\usepackage[T1]{fontenc} 
\usepackage{subfig}
\usepackage{slashed}
\usepackage{float}
\usepackage{amssymb}
\usepackage{rotating}
\usepackage{bm}
\usepackage{amsbsy}
\usepackage{lipsum}

\newcommand{\equaref}[1]{Eq.~(\ref{#1})}
\newcommand{\equasref}[2]{Eqs.~(\ref{#1})~and~(\ref{#2})}

\newcommand{\figref}[1]{Fig.~\ref{#1}}

\newcommand{\secref}[1]{Section~\ref{#1}}
\newcommand{\appref}[1]{Appendix~\ref{#1}}
\newcommand{\tabref}[1]{Table~\ref{#1}}

\newcommand{\tphi}{\tilde{\varphi}}
\newcommand{\tih}{\tilde{h}}
\newcommand{\tPhi}{\tilde{\Phi}}
\newcommand{\bq}{\begin{eqnarray}}
\newcommand{\nq}{\end{eqnarray}}
\newcommand{\Real}{\text{Re}}
\newcommand{\Imag}{\text{Im}}
\usepackage{accents}
\newlength{\dhatheight}
\newcommand{\doublehat}[1]{%
    \settoheight{\dhatheight}{\ensuremath{\hat{#1}}}%
    \addtolength{\dhatheight}{-0.35ex}%
    \hat{\vphantom{\rule{1pt}{\dhatheight}}%
    \smash{\hat{#1}}}}
\usepackage{xspace,amsmath,booktabs,graphicx}

\renewcommand{\vec}[1]{\ensuremath{\mathbf{#1}}\xspace}

\newcommand{\rivet}{\textsc{Rivet}\xspace}
\newcommand{\sherpa}{\textsc{Sherpa}\xspace}

\newcommand{\atlas}{\textsc{Atlas}\xspace}
\newcommand{\gmt}{\ensuremath{g-2}\xspace}
\newcommand{\meg}{\textsc{MEG}\xspace}

\newcommand{\MeV}{\text{Me\kern -0.15ex V}\xspace}
\newcommand{\GeV}{\text{Ge\kern -0.15ex V}\xspace}
\newcommand{\TeV}{\text{Te\kern -0.15ex V}\xspace}

\newcommand{\lgvphi}{\ensuremath{\log_{10}(v_\varphi)}\xspace}
\newcommand{\eps}{\ensuremath{\varepsilon}\xspace}
\newcommand{\lgeps}{\ensuremath{\log_{10}(\varepsilon)}\xspace}
\newcommand{\epsphi}{\ensuremath{\epsilon_{\varphi}}\xspace}

\newcommand{\gone}{\ensuremath{g_1}\xspace}

\newcommand{\lggone}{\ensuremath{\log_{10}(g_1)}\xspace}
\newcommand{\lggtwo}{\ensuremath{\log_{10}\left(-g_2\right)}\xspace}
\newcommand{\epmag}{\ensuremath{\left |\epsilon_{\varphi}\right|}\xspace}
\newcommand{\lgepmag}{\ensuremath{\log_{10}(\left|\epsilon_{\varphi}\right|)}\xspace}
\newcommand{\epphase}{\ensuremath{\theta_{\varphi}}\xspace}

\newcommand{\Mtwo}{\ensuremath{M^2_{\tPhi}}\xspace}
\newcommand{\cls}{\ensuremath{CL_s}\xspace}

\usepackage{array}
\usepackage{siunitx}
\usepackage{diagbox}
\usepackage[splitrule]{footmisc}
\usepackage{caption}

\preprint{\begin{flushright} 
FERMILAB-PUB-18-549-T\\
               NSF ACI-1450310\\
PHY-1505463
\end{flushright} }

\title{\textbf Constraining $A_4$ Leptonic Flavour Model Parameters  at Colliders and Beyond}

\author[a]{Lukas Heinrich,}
\author[b]{Holger Schulz,}
\author[c]{Jessica Turner}
\author[d]{and Ye-Ling Zhou}

\affiliation[a]{Physics Department, New York University, New York, NY 10003, U.S.A.}
\affiliation[b]{Department of Physics, University of Cincinnati, Cincinnati, OH 45219, U.S.A.}
\affiliation[c]{Theoretical Physics Department, Fermi National Accelerator Laboratory, P.O. Box 500, Batavia, IL 60510, U.S.A.}
\affiliation[d]{School of Physics and Astronomy, University of Southampton,
Southampton, SO17 1BJ, U.K.}

\emailAdd{lh1132@nyu.edu}
\emailAdd{schulzhg@ucmail.uc.edu}
\emailAdd{jturner@fnal.gov}
\emailAdd{ye-ling.zhou@soton.ac.uk}

\abstract{
The observed pattern of mixing in the neutrino sector may be explained by the
presence of a non-Abelian, discrete flavour symmetry broken into residual
subgroups at low energies. Many flavour models require the presence of
Standard Model singlet scalars  which can promptly decay to charged leptons
in a  flavour-violating manner.  We constrain
the model parameters of a generic $A_4$ leptonic flavour model using a synergy of experimental data including
limits from  charged lepton flavour conversion, an 8 TeV collider analysis and constraints from the anomalous magnetic moment of the muon.
 The
most powerful constraints derive from  the \meg collaborations' limit on
Br$\left(\mu\to e\gamma\right)$ and the reinterpretation of an 8 TeV \atlas search for
anomalous productions of multi-leptonic final states.  We quantify the exclusionary power of each of these experiments and identify
regions where the
constraints from collider and \meg experimental data are complementary.}

\begin{document} 
\maketitle
\flushbottom

\allowdisplaybreaks

\section{Introduction}\label{sec:intro}
Since the discovery of neutrino oscillations by Super-Kamiokande
\cite{Fukuda:1998mi},  two puzzling aspects of neutrino physics have emerged.
First, neutrinos have very small but non-zero masses and second,  the leptonic
mixing or Pontecorvo-Maki-Nakagawa-Sakata (PMNS) matrix, $U$, has  a strikingly
different structure from the quark mixing matrix.  One of the most fruitful  beyond the Standard Model (SM) ideas 
applied to the neutrino sector
is the introduction of a  non-Abelian flavour symmetry to explain the observed structure of the PMNS matrix.
 These models generally propose a discrete flavour symmetry  which is
broken spontaneously, leaving the leptonic mass terms invariant under residual symmetries. 
Through symmetry considerations alone, without the specification of a detailed flavour model, it is possible to reduce the number of degrees of freedom between mixing parameters and 
thereby predict sum-rules which will be testable at upcoming long  (T2HK and DUNE) \cite{Abe:2014oxa,Acciarri:2015uup} and medium (JUNO) \cite{An:2015jdp} baseline neutrino oscillation experiments. 

The popularity of the flavour symmetry paradigm is reflected in the sheer number of flavour symmetry groups
that have been considered: from continuous ones such as $U(1)$ \cite{Froggatt:1978nt}, $SO(3)$ \cite{Berger:2009tt, King:2018fke}, $SU(3)$ \cite{Joshipura:2009gi,Alonso:2013nca}, and also the  
discrete cases $Z_n$ \cite{Grimus:2004hf,Zhou:2012ds}, $A_4$ \cite{Ma:2001dn,Altarelli:2005yp,Altarelli:2005yx}, $A_5$ \cite{Ballett:2015wia,Li:2015jxa,DiIura:2015kfa}
$S_4$ \cite{Mohapatra:2003tw,Lam:2008rs}, 
$\Delta(27)$ \cite{deMedeirosVarzielas:2005qg,deMedeirosVarzielas:2006fc}, $\Delta(48)$ \cite{Ding:2013nsa,Ding:2014hva}, etc. For a comprehensive review see e.g., Refs \cite{Altarelli:2010gt,King:2013eh,King:2014nza}.  In a model-independent manner various leading order  mixing patterns emerge as a result of  flavour symmetries and their possible breaking such as tribimaximal
(TBM) \cite{Harrison:2002er,Xing:2002sw,Harrison:2002kp, He:2003rm}, golden-ratio (GR) \cite{Datta:2003qg,Feruglio:2011qq,Everett:2008et,Kajiyama:2007gx} and bimaximal (BM) mixing \cite{Fukugita:1998vn,Barger:1998ta,Davidson:1998bi,Altarelli:2009gn,Meloni:2011fx}.
In order to render the structure of the leptonic mixing compatible with data, in particular with the observation of a non-zero reactor
mixing angle $\theta_{13}\approx 8^{\circ}$ \cite{An:2012eh,An:2013uza,Ahn:2012nd,Ardellier:2006mn}, corrections to these mixing patterns are necessary.

Such a task can be completed in a model-independent or dependent manner. In the latter case, the breaking of the flavour symmetry is 
realised by SM singlet scalar fields, also known as \emph{flavons}, which have non-trivial quantum numbers associated to the non-Abelian flavour group.
These flavons acquire vacuum expectation values (VEVs) which spontaneously break the flavour symmetry to its Abelian residual symmetries in the charged lepton and neutrino sector. In general, two flavons are sufficient; however for larger symmetry groups and supersymmetric setups additional flavon multiplets are necessary for model construction
 \cite{Bazzocchi:2008ej,Bazzocchi:2009pv,Bazzocchi:2009da,Krishnan:2012me,Toorop:2010yh,Hagedorn:2010th,Meloni:2011fx,Zhao:2012wq,Ding:2014hva,Ding:2013nsa,Feruglio:2011qq,Gehrlein:2014wda,Gehrlein:2015dza,Ding:2012xx,Everett:2008et}.
Typically, the corrections to the leading order mixing pattern are provided by higher dimensional operators formed between the flavons and charged leptons
\cite{King:2013eh,Altarelli:2010gt,King:2014nza}. An alternative possibility was proposed in the work of \cite{Pascoli:2016eld}, where the cross-coupling between 
the flavons of the neutrino and charged lepton sector may slightly break the Abelian residual symmetries  and thereby provide the needed deviation from exact TBM mixing  in the context of an 
$A_4$ flavour model.  

The rich phenomenology of flavour models has been explored  in the quark and lepton sector using both Abelian and non-Abelian flavour symmetries.
 In the case of Abelian family symmetries, which manifests from the Froggatt-Nielsen (FN) mechanism \cite{Froggatt:1978nt},  the collider and flavour violating phenomenology of a single 
flavon was explored \cite{Tsumura:2009yf}.  Although our model and theirs markedly differ, we
reach a similar conclusion to their work: limits from MEG can largely exclude the flavour breaking scale of less than $\sim 1$ TeV.
In addition, there has been work completed on constraining quark flavour model parameter space using collider constraints including Higgs-flavon mixing, 
electroweak oblique parameters and direct production of the flavon at current and future colliders \cite{Berger:2014gga,ArroyoUrena:2018mvl,Bauer:2016rxs}.
Moreover, the observed flavour violating decay of the Higgs ($h\rightarrow \mu\tau$) was investigated in the context of 
 a FN mechanism \cite{Huitu:2016pwk}. 
 
Using non-Abelian discrete symmetries the CLFV processes in $A_4$ was first discussed in \cite{Ma:2010gs}, where channels allowed by the residual symmetry $\mathbb{Z}_3$ were emphasised. CLFV processes mediated by flavons were studied in \cite{Kobayashi:2015gwa, Pascoli:2016wlt}. Specifically, correlations between $\mathbb{Z}_3$-breaking channels and the correction to TBM were discovered in \cite{Pascoli:2016wlt}. Constraints on the flavon mass in  supersymmetric $A_4$ leptonic flavour models have been studied \cite{Muramatsu:2016bda}.
Moreover, the observed flavour violating decay of the Higgs ($h\rightarrow \mu\tau$) was investigated in the context of 
 $A_4$ \cite{Heeck:2014qea}. 
It was found that the flavon could be produced at colliders if it is sufficiently light. There is also prospect for direct production, without  reliance upon
the flavon-Higgs mixing, at lepton colliders \cite{Muramatsu:2017xmn}.  

The primary aim of this work is to  exclude regions  of
parameter space of a  non-supersymmetric $A_4$ leptonic flavour model. To do so  we apply  a synergy of experimental data ranging from the reinterpretation an 8 TeV collider analysis to applying limits from charged lepton flavour violating (CLFV) processes determined by the MEG collaboration. To our best knowledge we are the first to undertake such a rigorous investigation of a relatively generic leptonic flavour model \cite{Pascoli:2016wlt}.
We begin with a discussion of  the motivation for and the
basic principles underlying leptonic flavour models.  We further elucidate on
the specific model  in \secref{sec:modelandint} with a particular emphasis on
the relevant interactions for the 8 TeV ATLAS analysis and charged lepton
flavour violation limits we recast. In \secref{sec:confront} we discuss the model parameter space and sampling strategy. We first confront the model with experimental data from $g-2$ and MEG as detailed in  \secref{sec:gm2} and \secref{sec:mte} respectively. The
implementation of the Higgs-scalar mixing and Higgs width constraints  are presented in  \secref{sec:HiggsWM} and \secref{sec:HiggsWidth} respectively. The aforementioned constraints can be 
calculated analytically; however, excluding regions of the parameter space using a collider data reinterpretation is a more involved process and the tool-chain, ATLAS analysis and CLs method are
discussed at length in \secref{sec:collider}. Finally, we present our results and make concluding remarks in \secref{sec:results} and \secref{sec:summary}.

\section{The $A_4$ Leptonic Flavour Model} \label{sec:confrontdisable}
\subsection{Basic Mechanism}
The threefold repetition of fermion generations and their subsequent masses
and mixing structure, is arguably one of the most puzzling features of the SM.
One plausible explanation of the pattern of fermionic mixing is an underlying
flavour symmetry.  In regards to the lepton sector, non-Abelian, discrete
groups are a popular choice of family symmetry. This derives from the
observation that leptonic mixing is large and generically, before the reactor
mixing angle was measured, the entries of the PMNS matrix resembled
Clebsch-Gordan coefficients of discrete groups.

The basic premise of leptonic flavour models is that at sufficiently
high-energies there exists an underlying family symmetry, typically non-Abelian
and discrete, which unifies the three generations of leptonic doublets into a
single mathematical structure, such as a triplet of the flavour group. From the
observation of neutrino oscillations, it is clear that leptonic masses are
non-degenerate and therefore the non-Abelian flavour group cannot be a symmetry
of the low-energy effective Lagrangian.  As a consequence, it is assumed  that
the full flavour symmetry must be broken at low energies into two Abelian
residual symmetry groups which are unbroken in the charged-lepton and neutrino
sectors.  The realisation of this breaking manifests through the introduction
of new scalars, known as \emph{flavons} which are  usually assumed to be SM
gauge singlets. The  scalar potential of these flavons    is invariant under
transformations of the non-Abelian flavour symmetry at high-energies. However,
at the \emph{flavour breaking scale}, the non-trivial alignment of the VEVs of
the flavons spontaneously break the non-Abelian flavour symmetry to Abelian
residual symmetries in the neutrino and charged lepton sectors.  The forms of
the residual symmetries derive from consideration of the largest  possible
symmetry of each sector and  the structure inherited from the larger
non-Abelian flavour group. The most general discrete residual symmetry of the
charged lepton sector is a direct product of cyclic groups, $\mathbb{Z}_n$.  In
contrast, the largest symmetry of the complex, symmetric Majorana neutrino mass
matrix is  $\mathbb{Z}_2\times\mathbb{Z}_2$. However, it is possible a subgroup
thereof, namely $\mathbb{Z}_2$, could be a residual symmetry of the neutrino
mass matrix.  These remnant symmetries constrain the structure of the charged
lepton and neutrino mass matrices and thereby the structure of the leptonic
mixing matrix.
%%%%%%%%%%%%%%%%%%%%%%%%%%%%%
%%%%%%%%  MODEL AND INTERACTIONS      %%%%%%%
%%%%%%%%%%%%%%%%%%%%%%%%%%%%%
\subsection{$A_4$ Leptonic Flavour Models}\label{sec:modelandint} 
\begin{figure}[h]
\centering
\includegraphics[width=0.35\textwidth]{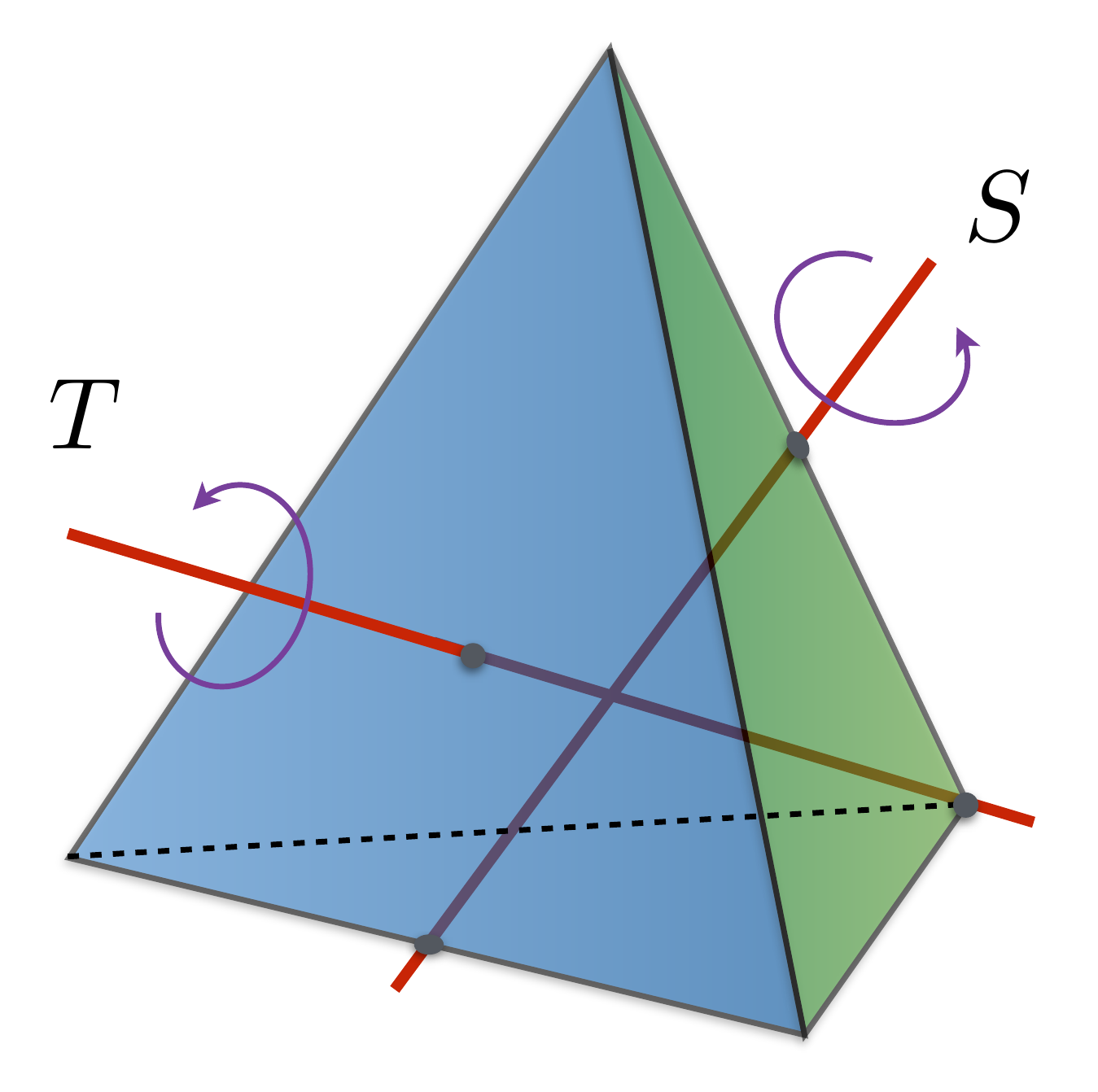}
\caption{The tetrahedral group $A_4$ as the full flavour symmetry and its subgroups $T=\mathbb{Z}_{3}$  and
$S=\mathbb{Z}_{2}$ as residual symmetries in the charged lepton and neutrino sectors, respectively.}
\label{fig:residual}
\end{figure}
In this work, we begin with the general $A_4$ model setup. The tetrahedral group  or the rotational
symmetries of a tetrahedron ($A_4$), as geometrically
represented in \figref{fig:residual},  is the smallest discrete group (order 12)
containing three-dimensional irreducible representations. 
The  generators  of the group, $T$ and $S$,  satisfy the relations: 
 $T^3=S^2=(ST)^3= \mathbb{1}$. 
  This group contains four irreducible representations: three singlets $\mathbf{1}$, $\mathbf{1}'$ $\mathbf{1}''$, and one triplet, $\mathbf{3}$. In the Altarelli-Feruglio basis, the triplet representation matrices of generators $S$ and $T$ are given by
\begin{equation}
T=\begin{pmatrix}
1 & 0 & 0\\
0 & \omega^2 & 0\\
0 & 0 & \omega
\end{pmatrix},
\quad
S=\frac{1}{3}\begin{pmatrix}
-1 & 2 & 2\\
2 & -1& 2\\
2 & 2 & -1
\end{pmatrix},
\end{equation}
where $\omega=e^{2i\pi/3}$. 
$T$ and $S$ are respectively the generator of the residual symmetries $\mathbb{Z}_{3} = \{\mathbb{1}, T, T^2\}$ and $\mathbb{Z}_{2} = \{\mathbb{1}, S\}$ after $A_4$ symmetry breaking. 
The only  physically inequivalent Abelian subgroups of $A_4$ are $\mathbb{Z}_3$ and $\mathbb{Z}_2$ where  all other Abelian subgroups are  conjugate to either $\mathbb{Z}_3$ or $\mathbb{Z}_2$. 

In flavour model building, at least two triplet flavons are required: one is needed for charged lepton and the other for neutrino mass generation.
 We denote these flavons as $\varphi$ and $\chi$, respectively.  These flavons could be a pseudo-real or a complex triplet of $A_4$.  In this present work, we focus on the former scenario where the three components of
$\varphi$ satisfy $\varphi_1=\varphi^*_1$ and $\varphi_2=\varphi^*_3$.
Such an assumption allows for the 
 minimal number of model parameters and degrees of freedom. 

In most $A_4$ models, the electroweak leptonic doublets (denoted as  $L=(L_{e},L_{\mu},L_{\tau})^T$ with $L_{e} = (\nu_{e L}, e_L)^T$, $L_{\mu} = (\nu_{\mu L}, \mu_L)^T$ and $L_{\tau} = (\nu_{\tau L}, \tau_L)^T$) are often arranged to belong to a $\mathbf{3}$ of $A_4$. 
And the right-handed charged leptons $e_R$, $\mu_R$ and $\tau_R$ belong to singlets $\mathbf{1}$, $\mathbf{1'}$ and $\mathbf{1''}$, respectively. The Higgs, $H$, is assigned as a trivial singlet $\mathbf{1}$ of $A_4$. At leading order, the general Lagrangian terms responsible for lepton masses have the following form 
\begin{eqnarray}
-\mathcal{L}_l &=& \frac{y_e}{\Lambda} (\overline{L} \varphi)_\mathbf{1} e_R H + \frac{y_\mu}{\Lambda} (\overline{L} \varphi)_{\mathbf{1}''} \mu_R H + \frac{y_\tau}{\Lambda} (\overline{L} \varphi)_{\mathbf{1}'} \tau_R H + \text{h.c.} \,, \nonumber\\
-\mathcal{L}_\nu &=& \frac{ y_1}{2\Lambda\Lambda_\text{W}} \big( (\overline{L} \tilde{H}\tilde{H}^T L^c)_{\mathbf{3}} \chi  \big)_\mathbf{1}  + \frac{y_2}{2\Lambda_\text{W}} (\overline{L} \tilde{H}\tilde{H}^T L^c)_\mathbf{1} + \text{h.c.} \,,
\label{eq:Yukawa_coupling} 
\end{eqnarray}
where $\tilde{H} = i \sigma_2 H^*$ and the subscript $\mathbf{r}$  ($\mathbf{r} = \mathbf{1}, \mathbf{1}', \mathbf{1}'', \mathbf{3}$) denotes the irreducible\footnote{We note that  a brief recap on the representation theory of $A_4$ can be found in \appref{sec:A_4reps}.} $\mathbf{r}$-plet product of the fields in the bracket. The scale $\Lambda$ is a new scale higher than $v_\varphi, v_\chi$ and 
may arise as a consequence of the decoupling of some heavy $A_4$ multiplet particles. In order to generate tiny Majorana neutrino masses, we apply the traditional dimension-five Weinberg operator $(\overline{L} \tilde{H}\tilde{H}^T L^c)$ where $\Lambda_\text{W}$ is the related scale, which may be different from $\Lambda$\footnote{$\mathcal{L}_\nu$ in Eq.~\eqref{eq:Yukawa_coupling} differs from that in \cite{Pascoli:2016eld} by the UV completion. In the latter case, the right-handed neutrino, as the UV completion, has been explicitly written out. The simplification in this paper does not influence the studies of flavon in the charged lepton sector. }.

The most widely studied mixing pattern is TBM mixing which  predicts $\sin\theta_{12}=1/\sqrt{3}$, $\sin\theta_{23}=1/\sqrt{2}$ and $\sin\theta_{13}=0$.
Naturally, corrections are required to render TBM mixing consistent with neutrino oscillation data, in particular with the non-zero valued reactor angle, $\theta_{13}\approx 8^\circ$.
 One great success of $A_4$ models is that they naturally predict TBM mixing (at leading order) based on the following symmetry argument. 

In order to ensure $\mathbb{Z}_3$ and $\mathbb{Z}_2$ as residual symmetries in charged lepton and neutrino sectors respectively, the vacuum alignment of these flavons is preserved under transformation of the residual
symmetries
\begin{equation}
T\langle \varphi \rangle = \langle \varphi \rangle\,, \quad  S\langle \chi\rangle = \langle \chi \rangle \,.
\end{equation}
As a consequence, VEVs of $\varphi$ and $\chi$ have to take the following forms
\begin{equation}
\langle  \varphi \rangle = 
 \begin{pmatrix}
1\\
0\\
0\\
\end{pmatrix}v_{\varphi} \,, \quad
\langle \chi \rangle = \begin{pmatrix}
1\\
1\\
1\\
\end{pmatrix}
\frac{v_{\chi}}{\sqrt{3}},
\end{equation}
respectively.
Substituting these VEVs into \equaref{eq:Yukawa_coupling}, in addition to the electroweak symmetry breaking VEV of the Higgs $\langle H \rangle = v_H / \sqrt{2}$ with $v_H=246$ GeV, we obtain the lepton mass matrices
\begin{eqnarray}
m_l=\left(
\begin{array}{ccc}
 y_e & 0 & 0 \\
 0 & y_\mu & 0 \\
 0 & 0 & y_\tau \\
\end{array}
\right)\frac{v_H v_\varphi}{\sqrt{2n}\Lambda} \,,\qquad
m_\nu=\left(
\begin{array}{ccc}
 2a+b & -a & -a \\
 -a & 2a & -a+b \\
 -a & -a+b & 2a \\
\end{array}
\right) \,,
\label{eq:lepton_mass}
\end{eqnarray}
where $a\equiv y_1v_\chi v_H^2/(4\sqrt{3}\Lambda\Lambda_\text{W})$ and $b \equiv y_2 v_H^2/(2\Lambda_\text{W})$. 
%At leading order, $M_l$ is diagonal and $M_\nu$ is diagonalised by the unitary matrix
%
We note that the  mass matrices of the
charged lepton ($m_{l}$) and neutrino ($m_{\nu}$) satisfy the aforementioned residual symmetries in the following
manner:
\begin{equation}
Tm_{l}m_{l}^\dagger T^\dagger = m_{l}m_{l}^\dagger,  \quad Sm_{\nu}S^T=m_{\nu},
\end{equation}
in which $T$ and $S$ are the 
generators of  $\mathbb{Z}_{3}$ and
$\mathbb{Z}_{2}$, respectively, as mentioned before. 
 
 In order to generate a leptonic mixing matrix consistent with current global
 fit data \cite{Esteban:2016qun}, there must be a slight breaking of either the
 residual symmetry of the neutrino or charged lepton sector or 
 possibly
 both. Although radiative corrections from the SM  break the exact TBM
 mixing,  such contributions are too small to induce $\lvert U_{e3}
 \rvert\approx 0.1$. 
It is possible that the necessary deviations from TBM  result from higher order operators in the flavon potential
or couplings between flavon and charged lepton sector (for reviews see e.g. \cite{Altarelli:2010gt, King:2013eh,King:2014nza}).
Such cross-couplings result in a VEV shift, i.e., $\langle \varphi_2 \rangle \neq 0$, which thereby breaks the residual  $\mathbb{Z}_3$ symmetry.
Without loss of generality, one can always perform the
following parameterisations
\begin{equation}
\langle \varphi_1 \rangle = v_\varphi\,, \quad 
\frac{\langle \varphi_2 \rangle}{\langle \varphi_1 \rangle} = \epsilon_\varphi\,.
\end{equation}
Using this parametrisation, the shifted VEV $\langle  \varphi \rangle$ can be always represented as  
 \begin{eqnarray}
\langle  \varphi \rangle = 
 \begin{pmatrix}
1\\
\epsilon_\varphi\\
\epsilon_\varphi^*\\
\end{pmatrix}v_{\varphi} \,,
 \end{eqnarray}
 where $\epsilon_\varphi$ is a complex parameter. 
 One requirement in this bottom-up approach is that, to be consistent with oscillation data, the $\mathbb{Z}_3$-breaking effect should be small. In particular, the shift $|\epsilon_\varphi| \ll 1$.  This shift could be one of the main sources of the deviations. In fact, as stated in \cite{Pascoli:2016eld}, if we assume all corrections to the mixing are obtained from the $\varphi$ VEV shift, $\theta_{13}$ and $\delta$ are predicted to be $\sin\theta_{13}=\sqrt{2}\text{Im} (\epsilon_\varphi)$ and $\delta=270^\circ -2 \text{Arg} (\epsilon_\varphi)$ for $\text{Im} (\epsilon_\varphi) >0$. 
 Furthermore, to be consistent with all oscillation data, $\epsilon_\varphi$ has to satisfy $0.10\lesssim|\epsilon_\varphi|\lesssim0.17$ and $38^\circ < \text{Arg} (\epsilon_\varphi) < 142^\circ$.
 However, in this work, we do not limit our discussion to a specific model. Instead, we will vary $\epsilon_\varphi$ in a relatively  wide range,  $|\epsilon_\varphi| \in [10^{-3}, 1]$ and $\text{Arg} (\epsilon_\varphi) \in [0, 2\pi)$ as shown in \tabref{tab:parambox}.  
 Such an approach allows us to be agnostic about the origin of the corrections to the mixing; 
 the needed correction could derive from a number of sources including 
 the shift in the VEV of 
 $\chi$ or  higher dimensional operators responsible for the lepton masses.

 \subsection{Interactions Relevant for Phenomenology}
We  study the observable
 phenomenology of this well motivated
 flavour model and therefore concentrate on the interactions of the flavon associated with the charged lepton sector ($\varphi$).
 For the flavon $\chi$ in the neutrino sector, it has lesser experimental visibility \footnote{Including non-standard interaction may lead to measurable effects of $\chi$ in neutrino oscillation experiments \cite{Wang:2018dwk}, but these effects are still small. }, which is why we do not consider its particle excitation and fix its VEV. 
 The $\varphi$ flavon communicates with the SM in two ways. The first is via
modification of the leptonic mass terms. The second is through the portal coupling of the flavons with the Higgs. 
In order to illustrate  the effective interactions involving flavons, we expand the flavons and  Higgs about their VEVs: 
\begin{eqnarray}\label{eq:excitation}
\varphi_i = \langle \varphi_i \rangle + \tilde{\varphi}_i \,, \quad
\text{Re}(H^0) = \frac{v_H + \tilde{h}}{\sqrt{2}} \,.
\end{eqnarray}

For the charged-lepton-portal interaction, we can straightforwardly 
 write  the couplings between flavons and charged leptons from the Lagrangian terms of \equaref{eq:Yukawa_coupling} in the Altarelli-Feruglio basis
     \begin{equation}\label{eq:CLFC}
    \begin{aligned}
    -\mathcal{L}_{\text{clfc}}^{\tih,\tphi_1} 
    &=\frac{m_e}{v_H} \overline{e} e \tih + \frac{m_\mu}{v_H} \overline{\mu} \mu \tih + \frac{m_\tau}{v_H} \overline{\tau} \tau \tih 
    + \frac{m_e}{v_\varphi} \overline{e} e \tphi_1 + \frac{m_\mu}{v_\varphi} \overline{\mu} \mu \tphi_1 + \frac{m_\tau}{v_\varphi} \overline{\tau} \tau \tphi_1\\
    &+ \frac{m_e}{v_H v_\varphi} \overline{e} e  \tphi_1  \tih + \frac{m_\mu}{v_H v_\varphi} \overline{\mu} \mu  \tphi_1  \tih + \frac{m_\tau}{v_H v_\varphi} \overline{\tau} \tau  \tphi_1  \tih\,, \\
    -\mathcal{L}_{\text{clfv}}^{\tphi_2}
    &= \frac{m_e}{v_\varphi} \big( \overline{\mu_L} e_R \tphi_2 +\overline{\tau_L} e_R \tphi_2^* \big) + \frac{m_e}{v_Hv_\varphi} \big( \overline{\mu_L} e_R  \tphi_2 +\overline{\tau_L} e_R  \tphi_2^* \big)  \tilde{h}\\
    &+ \frac{m_\mu}{v_\varphi} \big( \overline{\tau_L} \mu_R \tphi_2 +\overline{e_L} \mu_R \tphi_2^* \big)
    + \frac{m_\mu}{v_Hv_\varphi} \big( \overline{\tau_L} \mu_R  \tphi_2 +\overline{e_L} \mu_R   \tphi_2^* \big) \tilde{h}\\
    &+ \frac{m_\tau}{v_\varphi} \big( \overline{e_L} \tau_R \tphi_2 +\overline{\mu_L} \tau_R \tphi_2^* \big)
    +\frac{m_\tau}{v_Hv_\varphi} \big( \overline{e_L} \tau_R  \tphi_2 +\overline{\mu_L} \tau_R  \tphi_2^* \big) \tilde{h} + \text{h.c.}\,.
    \end{aligned}
    \end{equation}
   From
   \equaref{eq:CLFC}, we observe that $\tih$ and $\tilde{\varphi}_1$ partake in
   charged lepton flavour conserving (CLFC) interactions while $\tilde{\varphi}_2$ partakes in  the CLFV interactions. 
   
The Higgs-portal interaction is obtained from the scalar potential. In principle, the full scalar potential should include self-couplings of $H$, $\varphi$ and $\chi$, as well as their cross-couplings.
However, as we ignore the excitation of $\chi$, the scalar potential can be simplified and effectively represented as a potential involving only the self- and cross-couplings of $H$ and $\varphi$, and the VEV of $\chi$ contributes as a correction to the potential.

The self-couplings of Higgs are identical to the SM Higgs potential, given by
\begin{equation}\label{eq:higgspotential}
V_{\text{self}}(H)= \mu^2_H H^\dag H + \lambda (H^\dag H)^2,
\end{equation} 
where the minimum of this potential is achieved by  $\mu_H^2<0$ and
$\lambda>0$. In the unitary gauge, the Higgs doublet takes the form, $\langle H
\rangle = (0, v_H/\sqrt{2})^T$. 
The cross-coupling between $\chi$ and $H$, $H^\dag H (\chi \chi)_{\mathbf{1}}$, only corrects the quadratic coupling of the Higgs after $\chi$ acquires a VEV. Such a term can be absorbed by the redefinition of the parameter $\mu_H$ and need not be written out explicitly. 

The flavon can communicate with the visible sector via the Higgs
portal coupling which cannot be forbidden by imposing any symmetries. The only renormalisable $A_4$-invariant operator is $(H^\dag H) (\varphi \varphi)_{\mathbf{1}}$.  This part of the   potential  is given by
\begin{equation}\label{eq:cross}
V_{\text{cross}}(H,\varphi) = \frac{1}{2}\epsilon H^\dag H (\varphi_1^2 + 2 |\varphi_2|^2),
\end{equation} 
where $\epsilon$ is a real parameter. 
As the Higgs is $A_4$-invariant, the cross-coupling does not alter the
$\varphi$ VEV direction.  Consequently,  this cross-coupling term does not
contribute to leptonic flavour mixing.  As we shall see later, this term  will
lead to mixing between the Higgs and flavon  and therefore plays an important
role for the flavon production at the Large Hadron Collider (LHC).

The self-couplings of $\varphi$ is the origin of the breaking of $A_4$ to $\mathbb{Z}_3$. To simplify the couplings, an additional $Z_2'$ symmetry (or a larger Abelian symmetry including the transformation $\varphi \to -\varphi$) is usually imposed. With these considerations in mind, the most general $A_4$- and $Z_2'$-invariant self-couplings of $\varphi$ is given by
\begin{equation}\label{eq:flavonpotential}
V_{\text{self}}(\varphi)= \frac{1}{2} \mu^2_\varphi I_{1\varphi} + \frac{g_1}{4} I_{1\varphi}^2 + \frac{g_2}{4} I_{2\varphi},
\end{equation}
where 
\begin{equation}
I_{1\varphi} = \varphi_1^2 + 2 |\varphi_2|^2 \,,\quad
 I_{2\varphi} = \frac{1}{3} \varphi_1^4 - \frac{2}{3} \varphi_1 (\varphi_2^3 + \varphi_2^{*3}) + |\varphi_2|^4.
\end{equation}
 In order to achieve a nontrivial and stable vacuum, the conditions
 $\mu_\varphi^2<0$ and $g_1+g_2/3>0$ are required and applied
 throughout this work. 

With the present terms of the Higgs and flavon potential (c.f. \equaref{eq:higgspotential} and \equaref{eq:flavonpotential}),
 after spontaneous flavour breaking the leptonic mixing matrix would have exact TBM structure. In order to achieve the necessary deviation
needed, the cross-coupling terms between charged lepton and neutrino flavons must be present. 
  The cross-couplings between the Higgs and $\chi$ can be absorbed by the
 redefinition of $\mu_H^2$ and therefore the only cross-coupling term left is the $\mathbb{Z}_3$-breaking one, $( \varphi \varphi )_{\mathbf{1}''} ( \chi \chi )_{\mathbf{1}'}$. 
This term is effectively represented as 
\begin{equation}\label{eq:Z3breaking}
V_{\mathbb{Z}_3\!\!\!\!/}(\varphi)= \frac{1}{2} A (  \varphi_2^2 + 2\varphi_1 \varphi_2^* ) +\text{h.c.},
\end{equation} 
where $A$ is a complex parameter with mass dimension two. The other cross-couplings are trivial. For example, $( \varphi \varphi )_{\mathbf{1}} ( \chi \chi )_{\mathbf{1}}$ with $\chi=\langle \chi \rangle$ can be absorbed by the redefinition of $\mu_\varphi$, and $( \varphi \varphi )_{\mathbf{3}} ( \chi \chi )_{\mathbf{3}} = 0$ for the $\mathbb{Z_2}$-preserving VEV $\langle \chi \rangle$.  

Hence, the effective potential is constructed from Equations \eqref{eq:higgspotential}, \eqref{eq:cross}, \eqref{eq:flavonpotential} and \eqref{eq:Z3breaking}:
\begin{equation}
V(H, \varphi) = V_{\text{self}}(H) + V_\text{cross} (H, \varphi) + V_{\text{self}}(\varphi) + V_{\mathbb{Z}_3\!\!\!\!/}(\varphi) \,.
\end{equation} 

After minimisation of the Higgs and flavon potential, these parameters satisfy the following condition
\begin{equation}\label{eq:Aeqs}
\begin{aligned}
&\mu_H^2 + \lambda v_H^2 + \frac{1}{2} \epsilon v_\varphi^2 (1+ 2 |\epsilon_\varphi|^2) = 0 \,,\\
&\mu_\varphi^2 + g_1 v_\varphi^2 (1 + 2|\epsilon_\varphi|^2) + \frac{1}{3} g_2 v_\varphi^2 [1 - \text{Re}(\epsilon_\varphi^3)] + \frac{1}{2} \epsilon v_H^2 + A \epsilon_\varphi^* + A^* \epsilon_\varphi = 0\,,\\
&\mu_\varphi^2 \epsilon_\varphi + g_1 v_\varphi^2 (1 + 2|\epsilon_\varphi|^2) \epsilon_\varphi + \frac{1}{2} g_2 v_\varphi^2 [- \epsilon_\varphi^{*2} + |\epsilon_\varphi|^2 \epsilon_\varphi] + \frac{1}{2} \epsilon \epsilon_\varphi v_H^2 + A + A^* \epsilon_\varphi^* = 0\,.
\end{aligned}
\end{equation}
We note that the shifted VEV, $\langle \varphi \rangle = (1,\epsilon_\varphi,\epsilon_\varphi^*)^T v_\varphi$, gives rise to non-zero $\theta_{13}$ and CP violation. The parameter $A$ may be determined from  \equaref{eq:Aeqs} in the following manner
\begin{equation}\label{eq:Aeq}
\begin{aligned}
A \epsilon_\varphi^* + A^* \epsilon_\varphi^{*2} + 2 \text{Re}(A^* \epsilon_\varphi) |\epsilon_\varphi|^2 &= \underbrace{-\frac{1}{2} g_2 v_\varphi^2 \epsilon_\varphi^{*3} + \frac{1}{3} g_2 v_\varphi^2 |\epsilon_\varphi|^2 \Big[ 1 - \text{Re}(\epsilon_\varphi^3) - \frac{3}{2} |\epsilon_\varphi|^2 \Big]}_{x} \\
A & = \frac{\left(\epsilon_{\varphi}^*\right)^2 x^*-\epsilon_{\varphi} \left(x+2 i \left|
   \epsilon_{\varphi}\right| ^2 \Im\left[x\right]\right)}{\left| \epsilon_{\varphi}\right| ^2
   \left(-\left|  \epsilon_{\varphi}\right|^2+{\epsilon_{\varphi}^*}^3+\epsilon_{\varphi}^3-1\right)} \,.
\end{aligned}
\end{equation}

We now consider the masses of the Higgs and flavons 
modified by the $\mathbb{Z}_3$-breaking coupling.  After the shifted VEV
$\langle \varphi \rangle = (1, \epsilon_\varphi,\epsilon_\varphi^*)^T
v_\varphi$ is included, the mixing between $\varphi_1$ and $\varphi_2$, as well
as the Higgs with $\varphi_2$, is predicted. We obtain the mass term for all
scalars in the basis $\tPhi = (\tih, \tphi_1, \sqrt{2}\Real(\tphi_2),
\sqrt{2}\Imag(\tphi_2))^T$
\bq
- \mathcal{L}_\text{scalar masses} = \frac{1}{2} \tPhi^\dag M^2_{\tPhi} \tPhi \,,
\nq
where the mass matrix $M^2_{\tPhi}$ is a real $4 \times 4$ symmetric matrix with the following entries
\bq\label{eq:massBig}
(M^2_{\tPhi})_{11} &=& 2\lambda v_H^2 \,,\nonumber\\
(M^2_{\tPhi})_{22} &=& 2 (\gone+\frac{g_2}{3}) v_\varphi^2 + \frac{1}{3} g_2 v_\varphi^2 \Real(\epsilon_\varphi^3) - 2 \Real(A\epsilon_\varphi^*) \,,\nonumber\\
(M^2_{\tPhi})_{33} &=& - \frac{1}{3} g_2 v_\varphi^2[ 1 - \Real(\epsilon_\varphi^3) ] + \frac{1}{2}g_2 v_\varphi^2 |\epsilon_\varphi|^2 - 2 \Real( A \epsilon_\varphi^* ) + \Real\left( - g_2 v_\varphi^2 (\epsilon_\varphi^* - \frac{1}{2} \epsilon_\varphi^2) + 2 g_1 v_\varphi^2 \epsilon_\varphi^2 + A^* \right) \,,\nonumber\\
(M^2_{\tPhi})_{44} &=& - \frac{1}{3} g_2 v_\varphi^2[ 1 - \Real(\epsilon_\varphi^3) ] + \frac{1}{2}g_2 v_\varphi^2 |\epsilon_\varphi|^2 - 2 \Real( A \epsilon_\varphi^* ) - \Real\left( - g_2 v_\varphi^2 (\epsilon_\varphi^* - \frac{1}{2} \epsilon_\varphi^2) + 2 g_1 v_\varphi^2 \epsilon_\varphi^2 + A^* \right) \,,\nonumber\\
(M^2_{\tPhi})_{12} &=& v_H v_\varphi \epsilon \,,\nonumber\\
(M^2_{\tPhi})_{13} &=& \sqrt{2} v_H v_\varphi \epsilon \Real(\epsilon_\varphi) \,,\nonumber\\
(M^2_{\tPhi})_{14} &=& \sqrt{2} v_H v_\varphi \epsilon \Imag(\epsilon_\varphi) \,,\nonumber\\
(M^2_{\tPhi})_{23} &=& \sqrt{2} \Real \left(2 g_1 v_\varphi^2 \epsilon_\varphi - \frac{1}{2} g_2 v_\varphi^2 \epsilon_\varphi^{*2} + A \right) \,, \nonumber\\
(M^2_{\tPhi})_{24} &=& \sqrt{2} \Imag \left(2 g_1 v_\varphi^2 \epsilon_\varphi - \frac{1}{2} g_2 v_\varphi^2 \epsilon_\varphi^{*2} + A \right) \,, \nonumber\\
(M^2_{\tPhi})_{34} &=& \Imag\left( - g_2 v_\varphi^2 (\epsilon_\varphi^* - \frac{1}{2} \epsilon_\varphi^2) + 2 g_1 v_\varphi^2 \epsilon_\varphi^2 + A^* \right),
\nq
and $\tilde\varphi_{1}$ ($\tilde\varphi_{2}$) 
   denotes the particle excitation around the VEV of $\varphi_{1}$ ($\varphi_{2}$), i.e., \equaref{eq:excitation}. 
Numerically, $M_{\tPhi}^2$ can be diagonalised by a real $4\times4$ orthogonal matrix $W$ as $W^T M_{\tPhi}^2 W = \text{diag}\{m_h^2, m_{s_1}^2, m_{s_2}^2, m_{s_3}^2 \}$. The SM Higgs is denoted as $h$ ($m_{h} = 125$ GeV)\footnote{Clearly, the SM quartic coupling is fixed once the mass matrix, $M$, is diagonalised and the $(1,1)$ entry is fixed to be the Higgs mass squared.} with the three other scalar mass eigenstates denoted as 
$s_1$, $s_2$, $s_3$.

We relate the gauge to the mass basis in the following way:
\begin{equation}
\begin{pmatrix}
\tilde{h}\\
\tilde{\varphi_{1}}\\
\sqrt{2}\text{Re}\left( \tilde{\varphi}_2\right)\\
\sqrt{2}\text{Im}\left( \tilde{\varphi}_2\right)
\end{pmatrix}
=
\begin{pmatrix}
W_{00} & W_{01}& W_{02} & W_{03}\\
W_{10} & W_{11}& W_{12} & W_{13}\\
W_{20} & W_{21}& W_{22} & W_{23}\\
W_{30} & W_{31}& W_{32} & W_{33}\\
\end{pmatrix}
\begin{pmatrix}
h\\
s_{1}\\
s_{2}\\
s_{3}\\
\end{pmatrix}.
\end{equation}
Before we proceed we will summarise the model parameter space  relevant for limit setting. From \equasref{eq:Aeq}{eq:massBig}, we observe 
this model contains the following parameters:  $\epsilon$, $\epsilon_{\varphi}$, $v_{\varphi}$, $g_1$, $g_2$. 
We note that all the parameters are real 
apart from $\epsilon_{\varphi}$ and therefore  there are six free parameters.
Some salient features of this model include:
\begin{itemize}
\item The parameter $\epsilon$  controls the cross-coupling of the flavons with SM Higgs. This parameter is of crucial importance 
for the Higgs-flavon mixing and therefore the
 production of flavons at colliders and will be further discussed in  \secref{sec:collider}. 

\item $\epsilon_{\varphi}$ parametrises the breaking of  $\mathbb{Z}_3$ in the charged lepton sector. 
As $\epsilon_{\varphi}\rightarrow 0$, the $\mathbb{Z}_3$-preserving limit is reached and TBM mixing is recovered. Therefore, 
$\varphi_{2}$ does not acquire a VEV and the only 
mixing that occurs is between  $h$ and $\varphi_{1}$. Subsequently, only CLFC interactions are present.  

\item The flavour breaking scale is parametrised by the VEV of  $\varphi$: $v_{\varphi}$. Moreover, the masses of the flavons are functions of $v_{\phi}$ and therefore the presence of these  flavons, at colliders or otherwise, will be increasingly suppressed as the flavour breaking scale becomes larger\footnote{Unfortunately for a $4\times4$ mass matrix there are no closed analytic form for the masses of $s_{1}$, $s_{2}$ and $s_{3}$; however the masses are a complicated function of all the parameters in $\vec{p}$ and are approximately linear in $v_{\varphi}$.}. 

\item  $M^2$ is  diagonalised via $M^2=W \hat{M}^2 W^T$, where $\hat{M}^2$ is a diagonal matrix and  $W$ is a real orthogonal matrix. In the case $|M^2_{ij}|\ll |M^2_{jj}-M^2_{ii}|$, the non-diagonal entries of $W$ are approximately given by
\begin{equation}
O_{ij}\approx \frac{M^2_{ij}}{M^2_{jj}-M^2_{ii}} \,.
\end{equation}
In the limit $M^2_{jj}\approx M^2_{ii}$, the mixing between the scalars becomes ill defined. 
In order to avoid this regime, we explore regions of the parameter space where the flavon masses are non-degenerate \footnote{We ensure the difference between the flavon masses is $\geq 10$ MeV.} and  the three flavon masses lie outside a $10$ GeV window of the Higgs mass as will be explained further in \secref{sec:collider}.

\item In the majority of the parameter space the flavons can promptly decay to two charged leptons both in a manner which is charged lepton flavour conserving and violating. Moreover, as the coupling of the flavons to the charged leptons is proportional to the charged lepton mass, the dominant decay channel of the flavons is tau-dominated. 

\item $g_{1}$ and $g_{2}$ are dimensionless couplings present in the $A_4 \times Z_2'$ invariant flavon potential as shown in \equaref{eq:flavonpotential}.
 Their role is most easily understood in the limit $\epsilon_{\varphi}\rightarrow0$ where an exact $\mathbb{Z}_3$ residual symmetry in the charged lepton sector is recovered. In such a framework, $\varphi_2$ does not acquire a VEV, and masses of $s_2$ and $s_3$ are obtained from the quadratic terms of the scalar potential, $m_{s_2} = m_{s_3} = \sqrt{ -\frac{1}{3}g_2 v_\varphi^2}$. However, $\varphi_1$, does acquire a VEV and therefore  mixes with the Higgs. Subsequently, the mass eigenstates of the Higgs and $\varphi_{1}$ are
 \begin{eqnarray}
m_h^2 &=& 2 \lambda v_H^2 + \left(\lambda v_H^2-(\gone+\frac{g_2}{3}) v_\varphi^2\right) \left(\sqrt{1+\Big(\frac{ \epsilon v_H v_\varphi}{\lambda v_H^2 - (\gone+\frac{g_2}{3}) v_\varphi^2}\Big)^2} - 1 \right) \,, \nonumber \\ 
m_{s_1}^2 &=& 2 (\gone+\frac{g_2}{3}) v_\varphi^2 + \left((\gone+\frac{g_2}{3}) v_\varphi^2-\lambda v_H^2\right) \left(\sqrt{1+\Big(\frac{ \epsilon v_H v_\varphi}{\lambda v_H^2 - (\gone+\frac{g_2}{3}) v_\varphi^2}\Big)^2} - 1 \right) \,.\nonumber \\
\end{eqnarray} 
 In the realistic regime we are interested in, namely $\epsilon_{\varphi}\neq0$, the masses of $m_{s_{1}}$ and $m_{s_{2}}$ ($m_{s_{3}}$) are proportional to $(\gone+g_2/3)$ and $g_2$ respectively\footnote{We note in the $Z_{3}$-preserving scenario $m_{s_2} = m_{s_3}$ but in the $Z_{3}$-breaking case there is a splitting in those masses.}, however the relation no longer has a closed analytic as the $4\times4$ mass matrix must be diagonalised numerically.
\end{itemize}

%%%%%%%%%%%%%%%%%%%%%%%%%%%%%
%%%%%%%%  CONSTRAINTS      %%%%%%%

%%%%%%%%%%%%%%%%%%%%%%%%%%%%%
\section{Confronting the Model with Experimental Data}\label{sec:confront}
In the previous section, we reviewed the pertinent features of the model and presented
its free parameters. The objective of this section is to evaluate the extent to which 
existing measurements are able to constrain the allowed parameter space.
In order to do so, we  compare predictions of the model with dedicated data from
the \gmt, \meg and  \atlas experiments.  The former two experimental limits can be directly
compared with analytic calculations and are discussed in \secref{sec:gm2} and \secref{sec:mte} respectively.
The comparison of the model prediction with collider data 
requires a rather involved tool-chain based on Monte-Carlo event generators and
analysis software. We discuss the signatures of this flavour model at the LHC, Monte-Carlo event generation, 
the \atlas measurement and statistical methodology used
 in \secref{sec:collider}. 

\subsection*{Parameter Space and Sampling}
To simplify the numerical treatment, we apply the polar form of the complex
parameter $\epsphi=\epmag\cdot e^{i\epphase}$ and define our parameter space in terms
of its magnitude and phase. In those cases where a model parameter spans
several orders of magnitude, the sampling is performed logarithmically. The parameter
sampling boundaries are given in \tabref{tab:parambox}.%
\begin{table}[!h]
    \centering
    \begin{tabular}{l|rr}
        \toprule[2pt]
        Parameter \vec{p} & $\min(p)$ & $\max(p)$ \\% sampling boundary & Upper sampling boundary \\
        \midrule
        \lgvphi &  1  & 3   \\
        \lgeps  & -3  & 0 \\
        \lggone & -4  & 0   \\
        \lggtwo & -4  & 0   \\
        \lgepmag & -3   & 0 \\
        \epphase & 0 & $2\pi$ \\
        \bottomrule[2pt]
    \end{tabular}
    \caption{Parameter sampling boundaries.}
    \label{tab:parambox}
\end{table}

\subsection*{Sampling Strategy}\label{sec:sampling}
The sampling is undertaken in a random uniform fashion. By doing so, we ensure that
samples from sub-spaces exhibit the same uniform structure as the global
sampling space. A sampled point is then used to construct the mass matrix, \Mtwo
(\equaref{eq:massBig}), which we diagonalise numerically. We \emph{reject} a sampled
point if any of the following conditions on the resulting scalar masses
$m_{s_{1}}, m_{s_{2}}, m_{s_{3}}$ is fulfilled:
\begin{enumerate}
    \item Any flavon mass is too light, i.e. $m_{s_{i}}<10$~GeV, $i=1,2,3$. \label{en:mlight}
    \item All flavon masses are $>1$~TeV. \label{en:mlarge}
    \item Any flavon mass is too close to the Higgs --- $\left|m(s_i)-m_H\right|<5$~GeV for  $i=1, 2, 3$. \label{en:mhiggs}
    \item Any flavon mass which is not the Higgs is close to being degenerate -- $\left|m(s_i)-m(s_j)\right|<100$~MeV for  $i, j=1, 2, 3$. \label{en:mdiff}
    \item $\lambda (\gone+\frac{g_2}{3}) < \frac{\eps}{4}$.\label{en:vacstab}
    \item $\gone+\frac{g_2}{3}<0$.\label{en:gsum}
\end{enumerate}

The conditions \ref{en:mlight} and \ref{en:mlarge} ensure that the theory is
well behaved and that the production cross-section of the new scalars is not
too small while the requirements \ref{en:mhiggs} and \ref{en:mdiff} protect
against being in the regime of resonant mixing. The final two constraints
\ref{en:vacstab}, \ref{en:gsum}  guarantee vacuum stability \cite{Englert:2013gz} and that the scalar
masses are positive.

All points sampled from the parameter space specified in \tabref{tab:parambox}
that pass all six prerequisites are further tested in terms of compatibility with
experimental data as detailed in the following. 
The total number of points
passing the aforementioned cuts is  $N_\text{tot}=8865$, which we found to be a sufficient number of points to cover the model
parameter space. For each type of data,
$d$, we count the number of points not excluded by it, $N_\text{pass}^{(d)}$, 
and calculate the exclusion power as
$\frac{N_\text{tot}-N_\text{pass}^{(d)}}{N_\text{tot}}$.  We summarise the exclusionary powers of the following experimental
data in \secref{sec:results}.

%%%%%%%%%%%%%%%%%%%%%%%%%%%%%
%%%%%%%%  g-2     %%%%%%%
%%%%%%%%%%%%%%%%%%%%%%%%%%%%%
\subsection{$g-2$}\label{sec:gm2}
\begin{figure}[t!]
\centering
\includegraphics[width=0.45\textwidth]{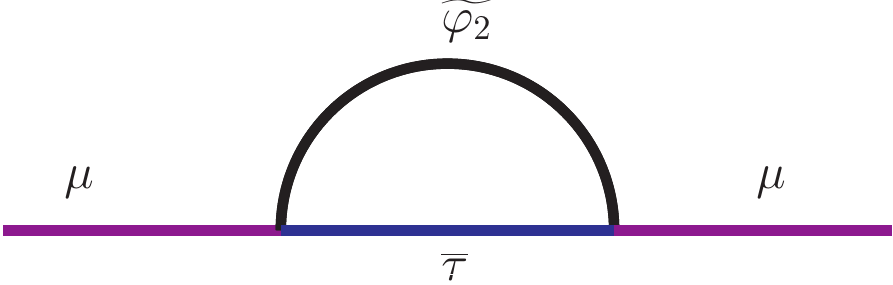}
\includegraphics[width=0.45\textwidth]{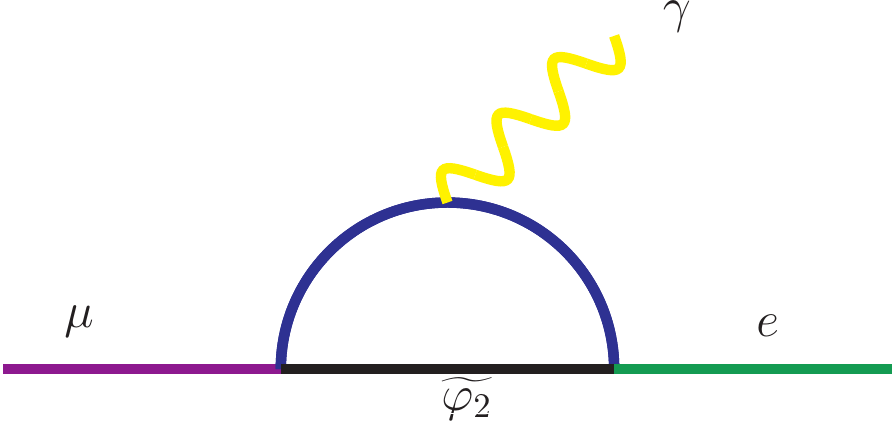}
\caption{On the left is the leading one-loop contribution to the muon anomalous magnetic moment and on the right is the leading one-loop contribution to $\mu\rightarrow e\gamma$. }
\label{fig:gm2}
\end{figure}
The most recent measurements of the anomalous magnetic moment of the
muon at Brookhaven
National Laboratory (BNL) \cite{Bennett:2006fi,Blum:2013xva} indicate a 
deviation from the
Standard Model precision calculation. The Muon $g-2$ experiment E989 based at Fermilab aims
to make a factor of four improvement upon the current measurement \cite{Grange:2015fou}.
 
 New physics models, with additional
scalars coupling to charged leptons, may explain this deviation.  The muon anomalous magnetic moment is
defined to be $a_{\mu} = \left( g-2\right)_\mu/2$ and its deviation from the SM
is given by
\bq
\begin{aligned}
\Delta a_{\mu}= a^{\text{exp}}_{\mu}-a^{\text{SM}}_{\mu}=\left(2.87 \pm 0.8 \right)\times 10^{-9} \,\left( 3.6\sigma \right).
\label{eq:gm2res}
\end{aligned}
\nq

In the flavour model we investigate, the leading contribution to $g-2$ has a
$\tau$-lepton running  in the loop as shown in \figref{fig:gm2}.  Completing a
standard calculation (see for example Refs. \cite{Queiroz:2014zfa,Keus:2017ioh,Lindner:2016bgg})
we find the magnetic moment to be
\bq
\begin{aligned}
\Delta a_{\mu}=\frac{m^2_{\mu}m^2_{\tau}}{24\pi^2v^2_{\varphi}}\Big[& \frac{ \left(\lvert W_{20}\rvert^2 - \lvert W_{30}\rvert^2\right) }{m^2_h}   + \frac{\left(\lvert W_{21}\rvert^2 - \lvert W_{31}\rvert^2\right)}{m^2_{s_{1}}} \\
+& \frac{\left(\lvert W_{22}\rvert^2 - \lvert W_{32}\rvert^2\right)}{m^2_{s_{2}}}  + \frac{\left(\lvert W_{23}\rvert^2 - \lvert W_{33}\rvert^2\right) }{m^2_{s_{3}}} \Big].
\label{eq:deltaamu}
\end{aligned}
\nq
We note that there is a one-loop level contribution to $\Delta a_{\mu}$ from
$\tilde{\varphi}_{1}$; however, as the couplings of each vertex $\propto
m_{\mu}/v_{\varphi}$, there is a $\mathcal{O}\left(10^{-3}\right)$ suppression
relative to that of the process shown in \figref{fig:gm2} and therefore its contribution is negligible.
In order to test whether a parameter point $\vec{p}$ is excluded by the result in \equaref{eq:gm2res},
we interpret the latter as an upper boundary on $a_{\mu}$ and  demand that $\Delta a_{\mu}(\vec{p}) \leq 3.68\times 10^{-9}$.

\subsection{\meg Result on $\text{Br}\left(\mu\rightarrow e \gamma\right)$}\label{sec:mte}

SM processes which violate charged lepton flavour, induced by massive neutrinos, occur at unobservable rates $\sim \mathcal{O}\left(10^{-48} \right)$.
However, new physics models which modify the charged lepton sector could enhance such processes to detectable rates and 
 provide crucial information in complement to direct searches. 
 The $\mathbb{Z}_3$-breaking flavour model discussed in \secref{sec:modelandint} has both  a rich flavour and chiral structure.  Moreover, as the flavons couple to the charged leptons, such a model will alter CLFV rates.
 
There has been a  systematic improvement in the sensitivity to a wide range  of CLFV processes.
The current bounds on the branching ratio of $\tau$-CLFV radiative decays processes are $\sim \mathcal{O}\left(10^{-8}\right)$ \cite{Aubert:2009ag,Hayasaka:2007vc,Hayasaka:2010np}.  The upper limit on the branching ratio of $\mu \rightarrow e \gamma $ flavour conversion processes are currently  $\sim \mathcal{O}\left(10^{-12}\right)$ \cite{Hayasaka:2010np, Bellgardt:1987du} with the most stringent constraint from the MEG collaboration, with  $ \text{Br}\left(\mu\rightarrow e\gamma\right)\leq 4.2\times10^{-13}$ \cite{TheMEG:2016wtm} at $90\%$ C.L. A MEG upgrade (MEG II) is envisaged to further constrain the upper limit on this CLFV process to $\sim 4\times 10^{-14}$ in the near term \cite{Ogawa:2015ucj}. The Mu2e experiment at Fermilab \cite{Bartoszek:2014mya} and COMET \cite{Kuno:2013mha} based in JPARC aims to even further increase the sensitivity to this rare decay, $\leq 10^{-16} - 10^{-17}$.

Consequently, as the experimental constraint from $\mu \rightarrow e \gamma $ flavour conversion provides one of the most  severe  limits on CLFV processes, the implications of this limit in both supersymmetric and non-supersymmetric $A_4$ flavour models from higher dimensional operators has has been studied in detail \cite{Feruglio:2008ht,Feruglio:2009hu}.  

For our scenario of   $\mathbb{Z}_3$-breaking scenario \footnote{We note there are a number of other 
CLFV transitions which may occur in this model such as $\tau\rightarrow \mu\gamma$, $\tau\rightarrow e\gamma$, $\tau\rightarrow e\mu\mu$ and $\tau\rightarrow e\mu\mu$ \cite{Pascoli:2016wlt}. However, as the limits placed on the branching ratio of these processes are relatively weak compared with the CLFV limit set by \meg, they will not offer stronger constraints.
} this process is loop-induced and is mediated by  $\tilde{\varphi}_{2}$ as shown in \figref{fig:gm2}. 
The contribution of the flavon in the loop was studied in great detail in \cite{Pascoli:2016wlt} where   the assumption of small $\epsilon_\varphi$ was applied. 
In this present work, we do not apply this assumption .

The leading contribution to this CLFV process, as  shown in \figref{fig:gm2},
    is mediated by $\varphi_{2}$.  Due to the flavour and chiral structure of
    the Yukawa couplings,  the dominant contribution derives from $\varphi_{2}$
    coupling to the $\tau$ charged leptons in the loop.  This contribution
towards $\text{Br}(\mu\rightarrow e \gamma)$ is calculated, in the mass basis,
and is given below:
%
%The leading contribution to this CLFV process, as  shown in \figref{fig:gm2}, is mediated by $\varphi_{2}$. We would anticipate 
%this as $\varphi_{2}$ (and its conjugate) has the only CLFV couplings c.f. \equaref{eq:CLFC}. More specifically the dominant contributions will 
%derive from $\varphi_{2}$ coupling to the $\tau$ charged leptons (as these couplings are proportional to $m_{\tau}/v_{\phi}$). 
%The Yukawa terms of \equaref{eq:CLFC} which couples $\varphi_{2}$ to $e$ and $\mu$ charged leptons do not 
%contribute to this particular radiative decay. This contribution of this process towards $\text{Br}(\mu\rightarrow e \gamma)$ 
%must be calculated in the mass basis as shown below:

%and present the analytic expression for the relevant bound on the branching ratio below
\bq
\text{Br}\left(\mu\rightarrow e\gamma \right) = \frac{\Gamma\left(\mu\rightarrow e\gamma \right)}{\Gamma\left(\mu\rightarrow e\overline{\nu_{e}}\nu_\mu \right)},
\nq
where 
\bq
\Gamma \left (\mu\rightarrow e\gamma \right) = \frac{m^3_{\mu} \lvert A \rvert^2}{16\pi}\,,
\quad
 \Gamma \left (\mu\rightarrow e\overline{\nu}_e\nu_{\mu} \right) = \frac{G_{F}^2m^5_{\mu}}{192\pi^3},
\nq
with
\begin{equation}
\begin{aligned}
A\left( h\right) &= \frac{1}{128\pi^2}\frac{1}{m^2_{h}v^2_{\varphi}} G_{2}\left( \frac{m^2_{\tau}}{m^2_H}\right)
\Bigg[  
m_{\mu}m^2_{\tau}\left( W_{20} + iW_{30}\right)^2 - m_{\mu}m^2_{\tau}\epsilon^*_{\varphi}
\left(\lvert W_{20} \rvert^2 + \lvert W_{30}\rvert^2 \right)
\Bigg],\\
A\left( s_{1}\right) &= \frac{1}{128\pi^2}\frac{1}{m^2_{s_{1}}v^2_{\varphi}}G_{2}\left( \frac{m^2_{\tau}}{m^2_H}\right)
\Bigg[  
m_{\mu}m^2_{\tau}\left( W_{21} + iW_{31}\right)^2 - m_{\mu}m^2_{\tau}\epsilon^*_{\varphi}
\left(\lvert W_{21} \rvert^2 + \lvert W_{31}\rvert^2 \right)
\Bigg],\\
A\left( s_{2}\right) &= \frac{1}{128\pi^2}\frac{1}{m^2_{s_{2}}v^2_{\varphi}}G_{2}\left( \frac{m^2_{\tau}}{m^2_H}\right)
\Bigg[  
m_{\mu}m^2_{\tau}\left( W_{22} + iW_{32}\right)^2 - m_{\mu}m^2_{\tau}\epsilon^*_{\varphi}
\left(\lvert W_{22} \rvert^2 + \lvert W_{32}\rvert^2 \right)
\Bigg],\\
A\left( s_{3}\right) &= \frac{1}{128\pi^2}\frac{1}{m^2_{s_{3}}v^2_{\varphi}}G_{2}\left( \frac{m^2_{\tau}}{m^2_H}\right)
\Bigg[  
m_{\mu}m^2_{\tau}\left( W_{23} + iW_{33}\right)^2 - m_{\mu}m^2_{\tau}\epsilon^*_{\varphi}
\left(\lvert W_{23} \rvert^2 + \lvert W_{33}\rvert^2 \right)
\Bigg].
\end{aligned}
\end{equation}

The functional form of $G_2$ is
\bq
G_2(x) = -\log x -\frac{11}{6},
\nq
and finally $A$ is given by the sum
\bq
A=A\left( h\right) + A\left( s_{1}\right) + A\left( s_{2}\right) + A\left( s_{3}\right).
\nq

Through the perturbative expansion of  $W_{ij}$, using $\epsilon_\varphi$ as
the expansion parameter, this result is consistent with that found in \cite{Pascoli:2016wlt}.
%Through the perturbative expansion of  $W_{ij}$, using $\epsilon_\varphi$ as the expansion parameter, this result is consistent with that found in \cite{Pascoli:2016wlt}.

The calculation is straightforward and we can compare the flavon model
prediction for $\text{Br}\left(\mu\rightarrow e\gamma \right)$ at any test
point $\vec{p}$ with the experimentally found upper limit.  We expect and
indeed find that MEG data provides the strongest exclusionary power of all the
experiments as it is dedicated to search for flavour change. This is discussed
in further detail in  \secref{sec:results}.

%The calculation is straight forward and we can compare the flavon model prediction for
%$\text{Br}\left(\mu\rightarrow e\gamma \right)$ at any test point $\vec{p}$ with the
%experimentally found upper limit. We find that the \meg data exhibits the strongest
%exclusionary power in this work. 
%%%%%%%%%%%%%%%%%%%%%%%%%%%%%
%%%%%%%%  Higgs width    %%%%%%%
%%%%%%%%%%%%%%%%%%%%%%%%%%%%%
\subsection{Higgs-Scalar Mixing Contraint}\label{sec:HiggsWM}
Extending the scalar sector of the SM has been a popular option to address
various beyond SM phenomena such as providing a dark matter candidate
\cite{Berlin:2015wwa,Kouvaris:2014uoa,Ghorbani:2014qpa}. Therefore,
implications on the Higgs sector in the context of a single pure  scalar
singlet have been explored in a number of works
\cite{Robens:2015gla,Pruna:2013bma,Godunov:2015nea,Falkowski:2015iwa,Dupuis:2016fda}.
These works have constrained the mixing of the new scalar with the Higgs. In
general, the mixing is small and the 125 GeV Higgs boson we observed at the LHC appears to
be mostly comprised of the SM Higgs mass eigenstate. 
In our case, there are three additional scalars which  acquire VEVs and
therefore  all three flavons  mix with the Higgs.  The constraint on $\vec{p}$ enforced
from mixing is imposed via the following requirement:
\[
\lvert W_{00} \rvert^2 > 0.86.
\]

\subsection{Higgs-Width Constraint}\label{sec:HiggsWidth}

In certain regions of the model parameter space, $\vec{p}$, it is possible that the
coupling of  $s_1$, $s_2$ and $s_3$ to $h$ will modify the Higgs total width.
Theoretical calculations, assuming purely SM interactions, predict the Higgs
total width ($\Gamma_{\text{SM}}$) to be $\approx4$ MeV. However, the
constrained upper limits of the total width, using measurements of on- and
off-shell decay rates to Z-bosons, indicate the upper limit to be $22$ MeV at
a 95$\%$ C.L. \cite{Khachatryan:2014iha}. In such regions of the model
parameter space, we assume that the deviation between the theoretically predicted
and measured Higgs width derives entirely from new physics associated  to our
model as  outlined  in  the Appendix\footnote{In the new physics contributions
to the Higgs width we ignore the three-body decays as these, in the majority of
the phase space, are  suppressed.}. We consider a point $\vec{p}$ excluded
by the Higgs-width results if the calculated width exceeds $22$~MeV.
 %%%%%%%%%%%%%%%%%%%%%%%%%%%%%
%%%%%%%%  COLLIDER     %%%%%%%
%%%%%%%%%%%%%%%%%%%%%%%%%%%%%
\subsection{Reinterpretation of \atlas Search for High Multiplicity Leptonic Final States}\label{sec:collider}

Reinterpreting a collider analysis is a more involved procedure compared to the
experimental constraints discussed in the previous sections. We shall therefore
initially give a brief explanation of the general workflow followed by more
specific descriptions of the \atlas analysis, event simulation and statistical
methods applied.

\subsubsection{General Workflow}

We need to simulate fully differential collider events and analyse them as
faithful to the original data analysis as possible in order to be able to
compare predictions to the measured observation and background events.  The procedure
requires writing the physics model in question in a language a Monte-Carlo (MC)
event generator is able to understand. We use FeynRules~\cite{Alloul:2013bka}
to code the Lagrangian and derive a model file in the Universal FeynRules
Output (UFO) format. This UFO format is understood by the MC event generator
\sherpa~\cite{Hoche:2014kca,Gleisberg:2008ta} which in turn is then able to simulate
proton-proton collisions according to the flavon model including QCD and QED
radiation effects as well as hadronisation and hadron decays.

The simulated events are analysed by the dedicated tool  \rivet~
\cite{Buckley:2010ar}. For the \atlas analysis we have chosen to reinterpret, we
are greatly helped by the fact that the analysis team provided a
validated \rivet routine of their measurement. 
The latter contains exactly the
same selection criteria and analysis logic as applied in the original data
analysis. One caveat, namely that the presented data and background
distributions have not been corrected for detector effects (not ``unfolded'') is
overcome by the fact that the analysis team included a machinery that applies
all resolutions and efficiencies to the simulated particles such that a fair
comparison between our signal MC and the data is possible.

In order to make a statement on whether a sampled point \vec{p} yields
a prediction that is compatible with the data we apply a hypothesis test
known as the \cls method. This method allows to distinguish, on a certain
confidence level, whether the observed data is more likely to be explained
by the SM background only or by the signal plus background hypothesis.

\subsubsection{The \atlas Search Analysis}

\captionsetup[subfigure]{labelformat=empty}
\begin{figure}[t]
    \subfloat[Signal region S1\label{sf:S1}]{%
        \includegraphics[width=0.48\textwidth]{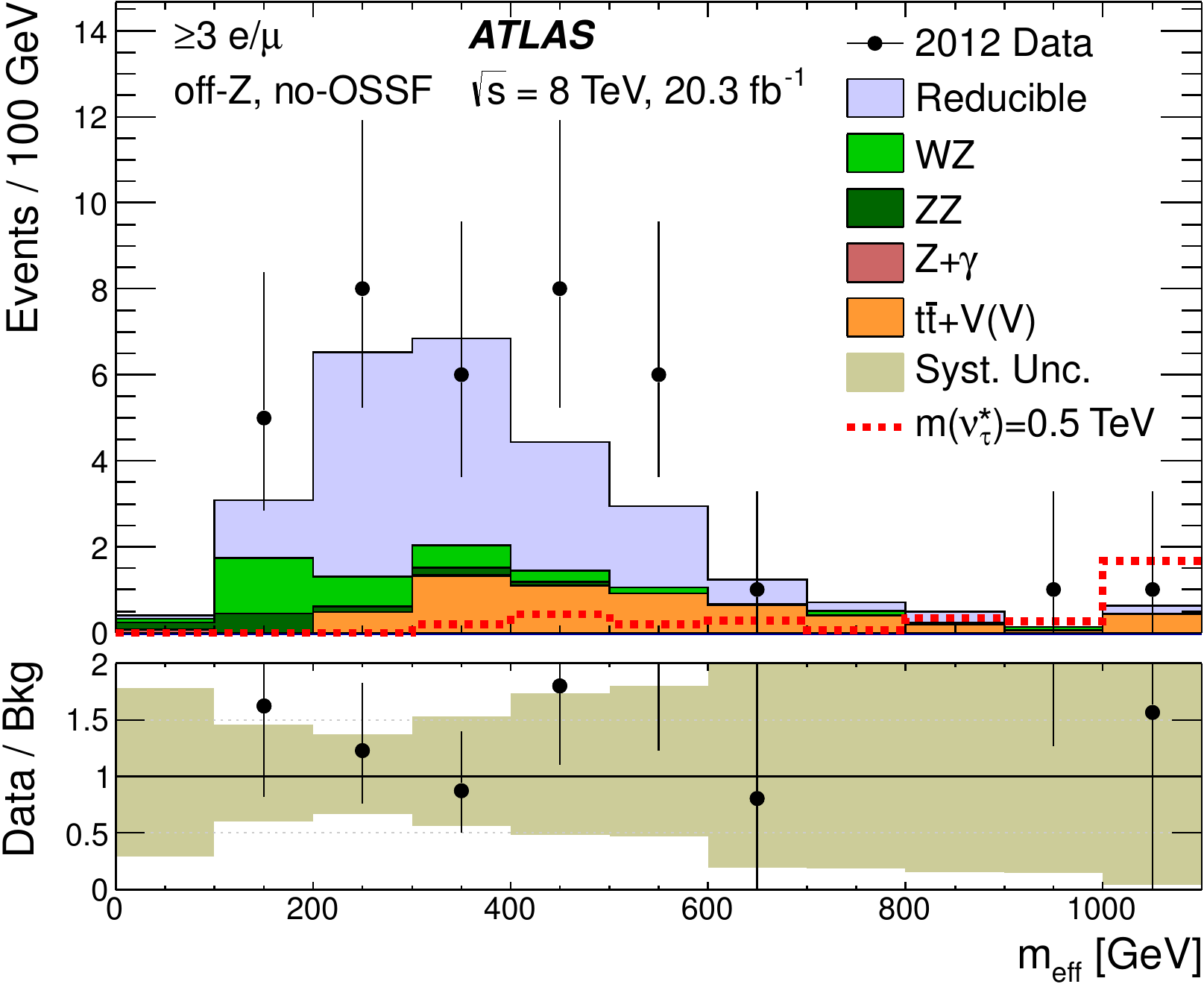}
     }
     \hfill
     \subfloat[Signal region S3\label{sf:S3}]{%
        \includegraphics[width=0.48\textwidth]{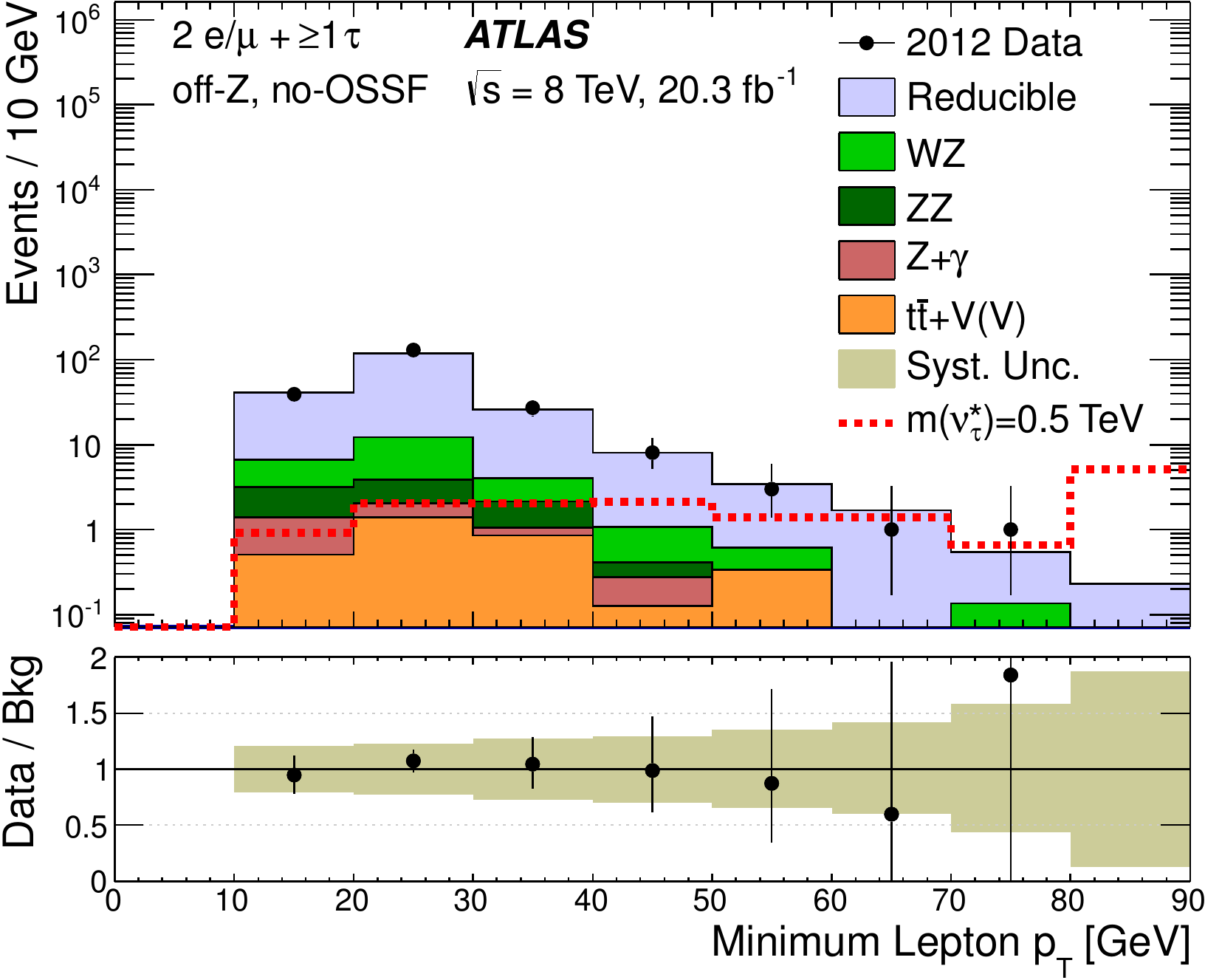}
     }\\
     \subfloat[Signal region S2\label{sf:S2}]{%
        \includegraphics[width=0.48\textwidth]{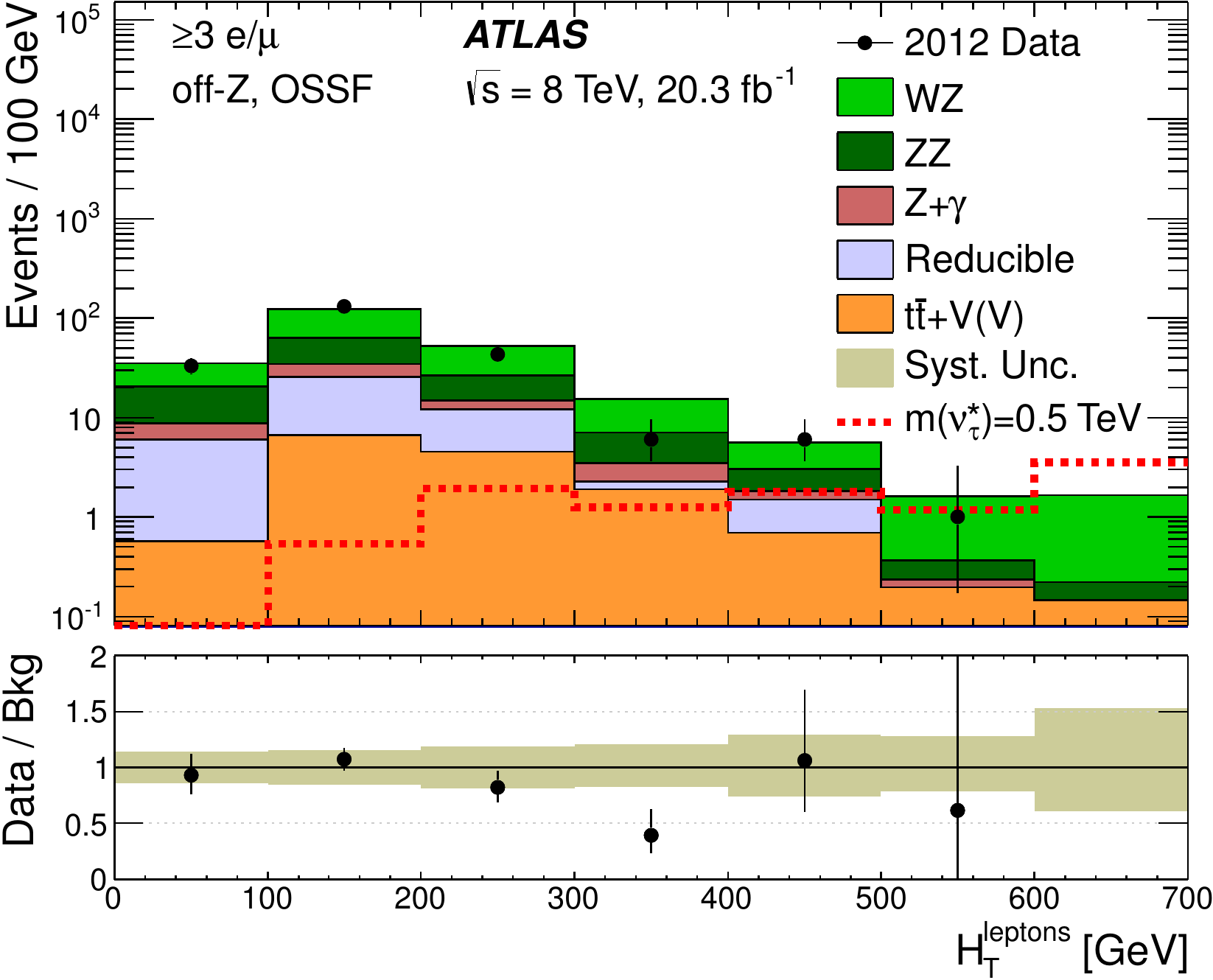}
     }
     \hfill
     \subfloat[Signal region S4\label{sf:S4}]{%
        \includegraphics[width=0.48\textwidth]{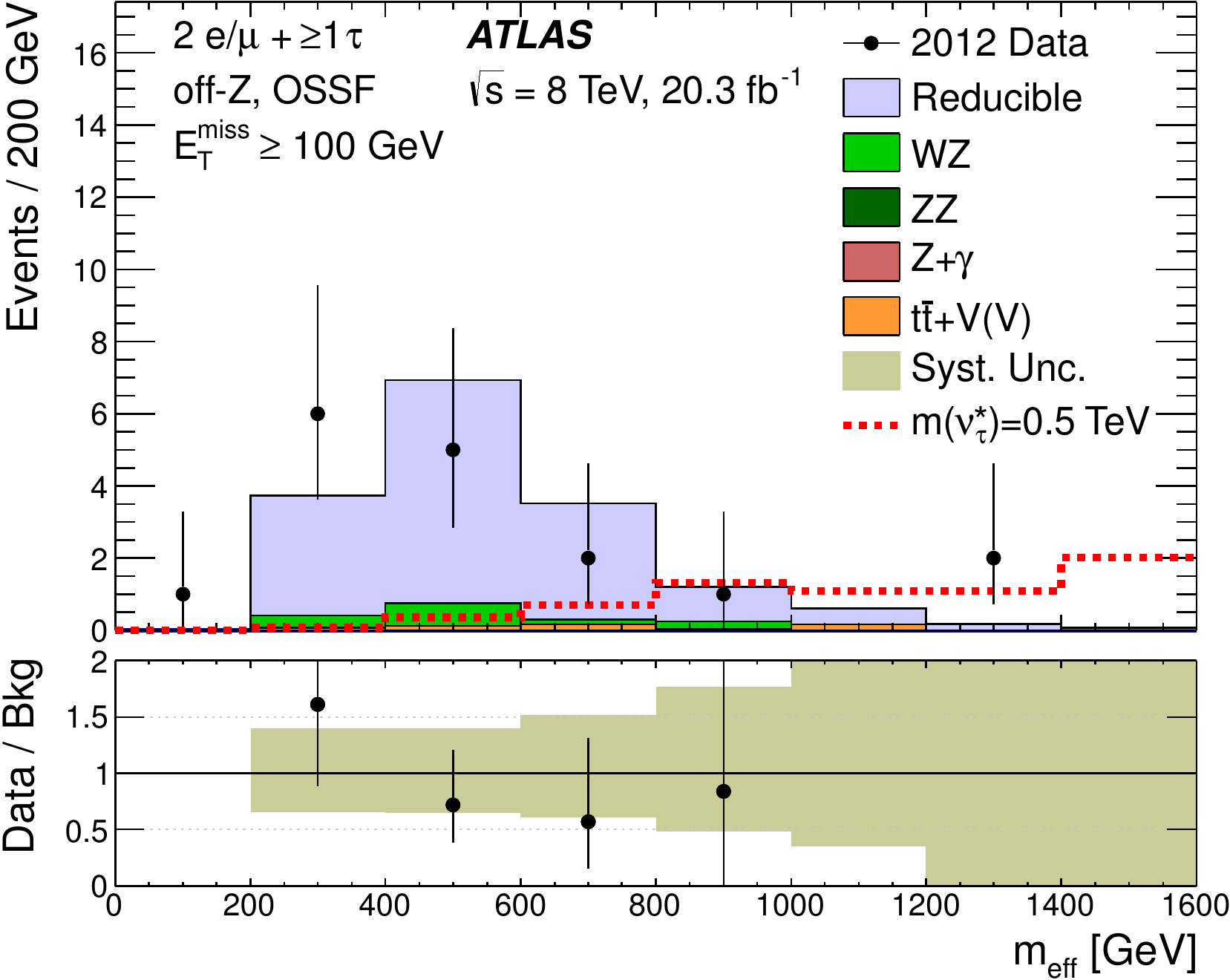}
     }
\caption{Data and background distributions of the \atlas search analysis~\cite{Aad:2014hja} used for reinterpretation in this work.
The definition of the signal regions S1 \ldots S4 is given in \tabref{tab:signalregions-def}. The data and background counts are explicitly listed in \tabref{tab:srcounts}. Copyright 2018 CERN for the benefit of the ATLAS Collaboration. CC-BY-4.0 license
}\label{fig:ATLASfigs}
\end{figure}

There are many beyond SM (BSM) scenarios which have anomalous production of
leptonic final states and therefore there have been a number of dedicated
analyses which have searched for three or more charged, prompt and isolated
leptons. These analyses have shown little deviation from the SM expectation and
therefore may be effective in excluding regions of parameter space for many
models. The ATLAS collaboration has conducted a number of supersymmetry
searches which have multi-lepton final states
\cite{Aad:2014nua,Aad:2014pda,Aad:2014iza} and indeed a model-independent
analysis was performed, providing limits using 7 TeV data \cite{Aad:2012xsa}.
The CMS Collaboration has also performed a similar analysis using both 7 TeV
\cite{Chatrchyan:2012mea} and 8 TeV data \cite{Chatrchyan:2014aea}. 

The analysis we choose to constrain our model parameter space, $\vec{p}$, uses
a data sample collected in 2012 by the ATLAS experiment with a centre of mass
energy of 8 TeV and  corresponding integrated luminosity of $20.3$ fb$^{-1}$
\cite{Aad:2014hja}.  This analysis  searches for the anomalous production of at
least three charged leptons in the final state. Moreover, as this analysis
searches for events which have at least one tau final state, it is
particularly well suited to our model as the flavons dominantly decay to taus
and muons. The analysis logic and data have been preserved, validated and made
publicly available by the \atlas collaboration within the analysis tool \rivet
\cite{Buckley:2010ar} as {\bf{ATLAS\_2014\_I1327229}}. 
We found this measurement, of all publicly available and validated analysis, to be 
the most suitable for constraining our model parameter space. We re-use the
observed data and total background estimates published by \atlas as presented
in \tabref{tab:srcounts} to perform our statistical analysis.

We shall not reiterate the full details of the analysis but rather present the
most relevant features for our work.
The analysis first applies a veto on Z-bosons and then divides the events
(which contain at least three leptons) into four disjoint signal regions based
on charged lepton flavour pairs and leptonic content.

\begin{itemize}
\item \textbf{OSSF}: events which contain an opposite sign same flavour (OSSF) charged lepton pair.
\item \textbf{no OSSF}: events which do not contain an OSSF pair.
\item \bm{$\geq 3e/\mu$}: events which contain minimally three electrons or muons.
\item \bm{$2e/\mu\geq 1\tau_{\textbf{had}}$}: events containing two electrons or muons with and at least one hadronically decaying tau lepton.
\end{itemize}

Depending on the signal region, different kinematic variables are used in the
measurement:
\begin{itemize}
\item $\pmb{H^{\text{leptons}}_{T}}$:  the scalar sum of $p_T$ of the leptons used to categorise the event.
\item \textbf{Minimum lepton} $\bm{p_T}$: minimum lepton transverse momentum.
\item $\pmb{m_{\text{eff}}}$: the effective mass of the event which combines the scalar sum of missing energy, scalar sum of the jets and the total $p_T$ of the leptons in the event.
\end{itemize}
An overview of the signal regions and observables applied is provided in \tabref{tab:signalregions-def} and \tabref{tab:signalregions-obs}.
The selection of histograms from \cite{Aad:2014hja}  we reinterpret in this work 
are shown in \figref{fig:ATLASfigs}\footnote{There are six histograms
in total, we choose the four which are most constraining for our model.}.  As
can be seen, the dominant SM processes which contribute to multi-leptonic final
states are diboson production and production of a top quark pair in association
with a weak gauge boson.  The statistics associated to the $m_{\text{eff}}$
kinematic variable are relatively low in comparison with the minimum lepton
$p_T$ and $H^{leptons}_{T}$.  Although, there is generally good agreement
between the SM predictions and the data; there are regions where the observed
event yield is lower than the expected yield by more than three times the
uncertainty on the expectation, and this occurs for the low statistics
histograms.  

\newcolumntype{x}[1]{>{\centering\let\newline\\\arraybackslash\hspace{0pt}}p{#1}}

\captionsetup[subfloat]{labelformat=empty}
\begin{table}[t]
    \centering
    \begin{subfloat}{}
        \begin{tabular}{r|x{3cm}|x{3cm}}
            %\toprule
                                 & \bm{$\geq 3e/\mu$} & \bm{$2e/\mu\geq 1\tau_{\textbf{had}}$}\\
            \midrule
                \textbf{no OSSF} &      S1            &    S3                                 \\
            \midrule
                \textbf{OSSF}    &      S2            &    S4                                 \\
        \end{tabular}
        \caption{Signal regions. Note that S4 has an additional missing transverse energy requirement of at least 100~GeV.}
        \label{tab:signalregions-def}
    \end{subfloat}
    \vspace{1cm}
    \begin{subfloat}{}
        \begin{tabular}{l|x{2.5cm}|x{2.5cm}|x{3.0cm}|x{2.5cm}}
            Signal region & S1 & S2 & S3 & S4 \\
            \midrule
            Observable    & $m_{\text{eff}}$ & $H^{\text{leptons}}_{T}$ & Min. lepton $p_T$ & $m_{\text{eff}}$  \\
        \end{tabular}
        \caption{Signal regions and observables used.}
        \label{tab:signalregions-obs}
    \end{subfloat}
    %\label{tab:signalregions}
\end{table}

\begin{table}
    \begin{minipage}{.48\textwidth}%
        \begin{center}
    % d02        
    \begin{tabular}{c|c|S[table-format=3.2]|S[table-format=3.2]|S[table-format=3.2]}
        \toprule[2pt]
        SR & Bin & $N_\text{obs}$ & $N_\text{BG}$ & $\Delta N_\text{BG}$\\
        \midrule
        \multirow{ 6}{*}{S1} & 2 & 5 & 3.08 & 0.43 \\
                             & 3 & 8 & 6.52 & 0.91 \\
                             & 4 & 6 & 6.84 & 0.71 \\
                             & 5 & 8 & 4.44 & 0.53 \\
                             & 6 & 6 & 2.95 & 0.39 \\
                             & 7 & 1 & 1.24 & 0.11 \\
                             & & & & \\
        %\midrule
        %\bottomrule[2pt]
    \end{tabular}\\
    % d01
    \begin{tabular}{c|c|S[table-format=3.2]|S[table-format=3.2]|S[table-format=3.2]}
        \toprule[2pt]
        SR & Bin & $N_\text{obs}$ & $N_\text{BG}$ & $\Delta N_\text{BG}$\\
        \midrule
        \multirow{ 6}{*}{S2} & 1 & 33  & 35.35  & 4.92  \\
                             & 2 & 132 & 123.04 & 16.46 \\
                             & 3 & 43  & 52.17  & 6.79  \\
                             & 4 & 6   & 15.23  & 1.36  \\
                             & 5 & 6   & 5.64   & 0.65  \\
                             & 6 & 1   & 1.17   & 0.09  \\
        %\midrule
        \bottomrule[2pt]
    \end{tabular}\\
        \end{center}
    \end{minipage}%
    \begin{minipage}{.04\textwidth}%
        \hfill
    \end{minipage}%
    \begin{minipage}{.48\textwidth}%
        \begin{center}
     %d04       
    \begin{tabular}{c|c|S[table-format=3.2]|S[table-format=3.2]|S[table-format=3.2]}
        \toprule[2pt]
        SR & Bin & $N_\text{obs}$ & $N_\text{BG}$ & $\Delta N_\text{BG}$\\
        \midrule
        \multirow{ 7}{*}{S3} & 2 & 39  & 41.13  & 8.40  \\
                             & 3 & 129 & 119.82 & 18.95 \\
                             & 4 & 27  & 25.89  & 3.59  \\
                             & 5 & 8   & 8.08   & 1.05  \\
                             & 6 & 3   & 3.44   & 0.41  \\
                             & 7 & 1   & 1.67   & 0.21  \\
                             & 8 & 1   & 0.54   & 0.07  \\
        %\midrule
        %\bottomrule[2pt]
    \end{tabular}\\
    %d03
    \begin{tabular}{c|c|S[table-format=3.2]|S[table-format=3.2]|S[table-format=3.2]}
        \toprule[2pt]
        SR & Bin & $N_\text{obs}$ & $N_\text{BG}$ & $\Delta N_\text{BG}$\\
        \midrule
        \multirow{ 6}{*}{S4} & 2 & 6 & 3.72 & 0.50 \\
                             & 3 & 5 & 6.93 & 0.87 \\
                             & 4 & 2 & 3.51 & 0.46 \\
                             & 5 & 1 & 1.19 & 0.15 \\
                             & 6 & 0 & 0.61 & 0.08 \\
                             & 7 & 2 & 0.17 & 0.03 \\
        %\midrule
        \bottomrule[2pt]
    \end{tabular}\\
        \end{center}
    \end{minipage}%
    \centering
    \caption{Observed ($N_\text{obs}$) and background ($N_\text{BG}$) counts reported by \atlas for each signal region (SR) used in this analysis .}
    \label{tab:srcounts}
\end{table}

\subsubsection{Event Simulation and Analysis}

The Universal FeynRules Output (UFO) for the model is generated using FeynRules
\cite{Alloul:2013bka} The model information in the Universal FeynRules Output
format is imported into the \sherpa event generator \cite{Hoche:2014kca} to
provide a full simulation of BSM processes at the particle level.

As the Higgs portal coupling, shown in \equaref{eq:cross}, is the only way the
flavons can be produced at a hadron collider, gluon fusion will be the most
relevant production channel for the flavons.  Although, BSM is available at
next-to-leading order (NLO) accuracy in NLOCT prepackaged in FeynRules
\cite{Degrande:2014vpa}, this feature is currently not implemented in Sherpa.
Therefore, we simply correct the tree-level
cross sections with a  K-factor of 2.2  for the BSM processes in this model \cite{deFlorian:2016spz}. This K-factor
is  computed at next-to-next-to leading order
\cite{Harlander:2002wh,Anastasiou:2002yz,Ravindran:2003um}, for  gluon fusion
which is the dominant production mechanism of the Higgs \cite{Georgi:1977gs}.
As the flavons decay leptonically, additional radiations from the final states
should not affect our results significantly. There are several kinematic
regimes which are important for flavon production:

\begin{itemize}
\item $m_{s_{i}}<\frac{m_{h}}{2}$ \\
In this kinematic regime, the flavons may be pair produced by the Higgs.
%\item $m_{h}<m_{s_{i}}<2m_{h}$\\
%The three body decay $s_{i}\rightarrow h f\overline{f}$ is kinematically
%allowed, however this will not only be suppressed by phase space but also the
%production will be suppressed by the mixing term.
\item $m_{s_{i}}<m_{h}$\\
The three-body decay $h_{i}\rightarrow s f\overline{f}$ is kinematically
allowed, however this will be suppressed by phase space.
\item $2m_{h}<m_{s_{i}}$\\
The pair production of the Higgs becomes kinematically accessible.
\end{itemize}

\begin{figure}[t]
\centering
\includegraphics[width=1.0\textwidth]{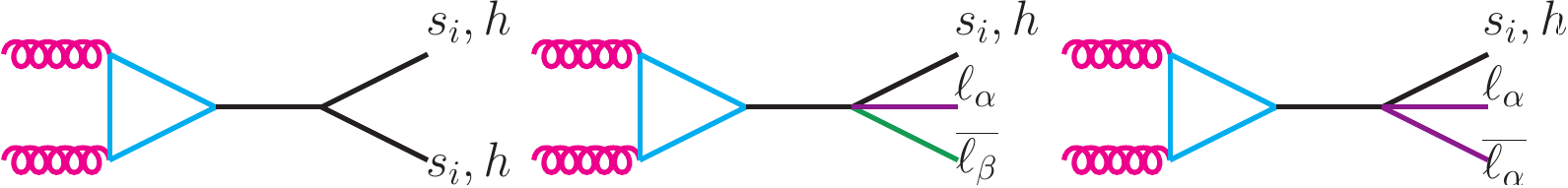}
\caption{Flavon production channels.}
\label{fig:colliderproduction}
\end{figure}

From \equaref{eq:CLFC}, we observe that  three-body decays of the scalars to scalar
and two leptons are possible but they are expected to be subdominant due to
phase space suppression. However, for certain points  in the model parameter
space, $\vec{p}$, it is possible that the three-body decays are non-negligible
\cite{Dupuis:2016fda} and therefore are included in our Monte-Carlo
simulations. The dominant configuration is two scalars in the final state where
all possible combinations are generated. The subdominant contribution is one
scalar in the final state with two leptons (both in  charged lepton flavour
conserving and violating combinations) as shown in \figref{fig:colliderproduction}. \sherpa
uses the matrix element generator Comix~\cite{Gleisberg:2008fv} to find all contributing
diagrams.

To summarise, for each test point \vec{p} we simulate $10^6$ events\footnote{The necessity 
for such a high number of generated signal events compared to the few hundreds of observed and background events comes from the relatively low
tau-efficiencies.} with \sherpa in a setup that includes the following
processes:
\begin{itemize}
    \item $pp\to s_i s_j$ where the flavons are denoted as $s_i$ and $s_j$ for $i, j=\{1, 2 ,3\}$.
        \item $pp\to s_i \ell\bar{\ell}$ where $s_i$ for $i=\{1, 2 ,3\}$ and $\ell$ denotes the charged leptons.
\end{itemize}
The generated scalars are decayed according to internally calculated
branching ratios. Furthermore, \sherpa takes care of QED and QCD radiation as
well as hadronisation and effects such as hadron decays and multiple parton interactions.
These events are then passed through the ATLAS analysis using \rivet and the
output is a signal histogram of expected event yields for each $\vec{p}$ that
can be compared with data and background estimates since \rivet applies all relevant
detector effects.

 %%%%%%%%%%%%%%%%%%%%%%%%%%%%%
%%%%%%%%  EVENT GENERATION      %%%%%%%
%%%%%%%%%%%%%%%%%%%%%%%%%%%%%
%\subsection{Event Generation}\label{sec:eventgen}
%%%%%%%%%%%%%%%%%%%%%%%%%%%%%
%%%%%%%%  CLs METHOD      %%%%%%%
%%%%%%%%%%%%%%%%%%%%%%%%%%%%%

\subsubsection{The CLs Method}\label{sec:CLs}

In order to infer information on the viability of a given point $\vec{p}$, the
expected event yield within the signal regions of the analysis is compared to
the observed data and estimated backgrounds as reported by the collaboration on
HepData \cite{hepdata_analysis}. As the experiment has not reported a
discrepancy between data and the estimated backgrounds, we set an upper limit
on the signal cross section normalised to the nominal cross section as
calculated by SHERPA through frequentist interval estimation based on the
\emph{profile likelihood} \cite{Cowan:2010js}. In this model, the probability of
observing data $\mathcal{D}$ is a function of parameters that are grouped into
two sets: the \emph{parameters of interest} (POIs), in this case the normalised
cross section $\mu$, as well as \emph{nuisance parameters} $\theta$.
The log-likelihood ratio is given by

\begin{equation}
    \lambda(\mu) = \frac{p(\mu, \doublehat{\theta} | \mathcal{D})}{p(\hat{\mu},\hat{\theta}|\mathcal{D})},
\end{equation}
where $\hat\mu$ and $\hat\theta$ are the best-fit value for an unconstrained
fit of the model against the data and $\doublehat\theta$ are the best-fit
values for the nuisance parameters for a constrained fit with a constant signal
strength $\mu$.
We choose to use the reported per-bin uncertainty as a shape-systematic on the
reported estimated background, such that the model has one nuisance parameter
$\gamma_i$ for each bin entering the fit.

In order to set an upper limit, we use the test statistic $q_\mu$
\begin{equation}
    q_\mu = \begin{cases}
              -2\log\lambda(\mu)\;\text{for}\; \hat\mu \leq \mu\\
              0\;\text{for}\; \hat\mu > \mu,
              \end{cases}
\end{equation}
where the choice of test statistic is only dependent on the parameter of interest
and avoids counting upward fluctuations in which the best fit value $\hat\mu$
exceeds the tested signal strength $\mu$ as evidence against signal hypothesis.

For the hypothesis test, we evaluate the modified $p$-value $\mathrm{CL}_s$ which is
commonly used by  collider experimentalists and is defined to be

\begin{equation}
  \mathrm{CL_s}(\mu) = \frac{\mathrm{CL_{s+b}}}{\mathrm{CL}_b} = \frac{\int_{q_{\mu,\mathrm{obs}}}^\infty p(q_\mu|\mu' = \mu)\mathrm{d}q_\mu}{\int_{q_{\mu,\mathrm{obs}}}^\infty p(q_\mu|\mu' = 0)\mathrm{d}q_\mu},
\end{equation}
in which $p(q_\mu|\mu')$ is the distribution of the test statistic $q_\mu$ for
data, which is populated according to a signal strengh $\mu'$ which we compute
using the asymptotic formulae derived in \cite{Cowan:2010js}. To compute the
$\mathrm{CL}_s$ values, the Python-based implementation of
HistFactory \cite{Cranmer:1456844} \verb+pyhf+ was applied \cite{pyhf}.

Subsequently, generated signal points are assessed at nominal signal strength $\mathrm{CL_s}
= \mathrm{CL_s}(\mu = 0)$ and points for which $\mathrm{CL}_s < 0.05$ are
considered to be excluded at 95\% confidence level.
%%%%%%%%%%%%%%%%%%%%%%%%%%%%%

\section{Results}\label{sec:results}

%%%%%%%%%%%%%%%%%%%%%%%%%%%%%
\begin{table}[t!]
    \centering
    \begin{tabular}{l|r}
        \toprule[2pt]
        Experimental data & Exclusion power[\%] \\% sampling boundary & Upper sampling boundary \\
        \midrule
        \meg (\secref{sec:mte}) & 65.6   \\
        \atlas (\secref{sec:collider}) &  40.0 \\
        Higgs-width (\secref{sec:HiggsWidth}) &  6.0   \\
        Higgs-mixing (\secref{sec:HiggsWM})&  1.7   \\
        \gmt (\secref{sec:gm2}) & 0.7 \\
        \bottomrule[2pt]
    \end{tabular}
    \caption{Exclusion power of constraints derived from experimental data.}
    \label{tab:exclusionfraction}
\end{table}
%%%%%%%%%%%%%%%%%%%%%%%%%%%%%

\begin{table}[t!]
%\begin{table}
    \centering
    \begin{tabular}{lc|c|c|c|c|c}
        \toprule[2pt]
                             \diagbox[width=18em,height=3em]{Constraint $j$}{Constraint $i$} &  & \meg & \atlas & H-width & H-mixing & \gmt \\
                             \midrule[2pt]
        \multicolumn{2}{c}{Number of points {\bf not} excluded by $i$ \quad} & 3045 & 5317 & 8331 & 8710 & 8806 \\
        \midrule[1.5pt]
        && \multicolumn{5}{c}{ of those, $j$ excludes}\\
        \midrule
                                 \meg    & & ---  & 2386 & 5469 & 5746 & 5761 \\
                                 \atlas  & & 114 & ---   & 3164 & 3452 & 3489 \\
                                 H-width & & 183  & 150  & ---  & 490  & 523  \\
                                 H-mixing& & 81   & 59   & 111  & ---  & 155  \\
                                 \gmt    & & 0    &  0   & 48   & 59   & ---  \\
                                 \midrule[2pt]
        \multicolumn{2}{c}{Number of points  excluded by $i$} & 5820 & 3707 & 534 & 155 & 59 \\
        \midrule[1.5pt]
        && \multicolumn{5}{c}{ of those, $j$ excludes}\\
        \midrule
                                 \meg    & & ---  & 3434 & 351 & 74  & 59  \\
                                 \atlas  & & 3434 & ---  & 384 & 96  & 59  \\
                                 H-width & & 351  & 384  & --- & 44  & 11  \\
                                 H-mixing& & 74   & 96   &  44 & --- & 0   \\
                                 \gmt    & & 59   &  59  &  11 & 0   & --- \\
        \bottomrule[2pt]
    \end{tabular}
    \caption{Summary table of exclusionary power for all 8865 points analysed. The table on top demonstrates the complementarity of e.g. \meg and \atlas --- of the 3045 points \emph{not} excluded by \meg, the \atlas data is able to exclude 114 points while \meg is able to exclude 2386 of the 5317 points \emph{not} excluded by the \atlas data. Similarly, the bottom table shows the overlap of exclusion of e.g. \meg and \atlas --- of the 5820 points excluded by \meg, the \atlas data is able to exclude 3434 points. }
    \label{tab:crossexclusion}
\end{table}

%%%%%%%%%%%%%%%%%%%%%%%%%%%%%
\begin{figure}[ht]
    \centering
    \includegraphics[width=\textwidth]{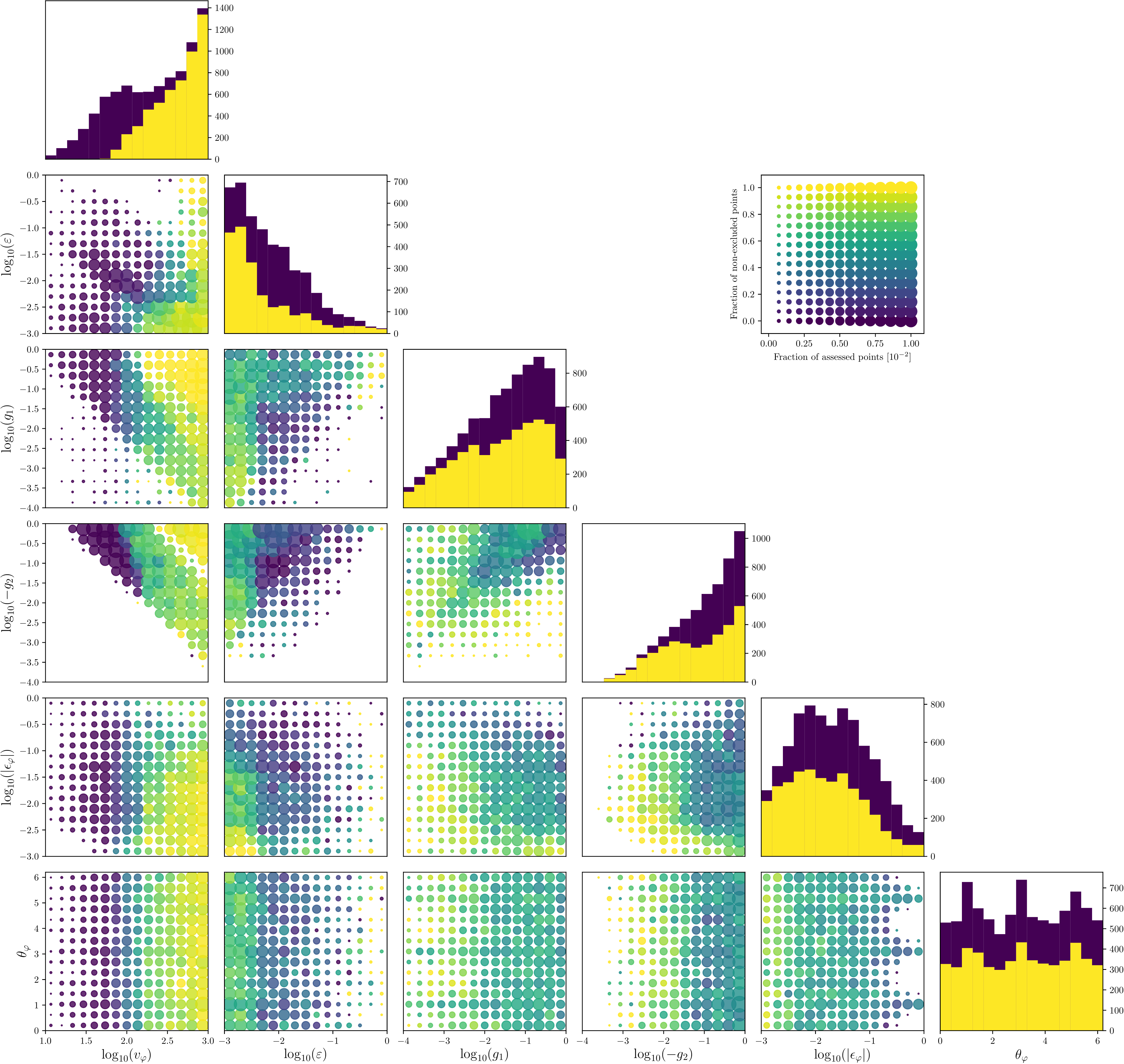}  
    \caption{Results for \atlas only. The histograms show the 1-dimensional projections of the number of excluded (purple) and not excluded (yellow) points. The scatter plots are a representation of 2-dimensional projections. The size of the circle indicates the fraction of the $N_\text{tot}$ points analysed in a single bin while the colour shows what fraction of those points can be considered excluded.}
      \label{fig:resultsATLAS}
\end{figure}
\begin{figure}[ht]
    \centering
    \includegraphics[width=\textwidth]{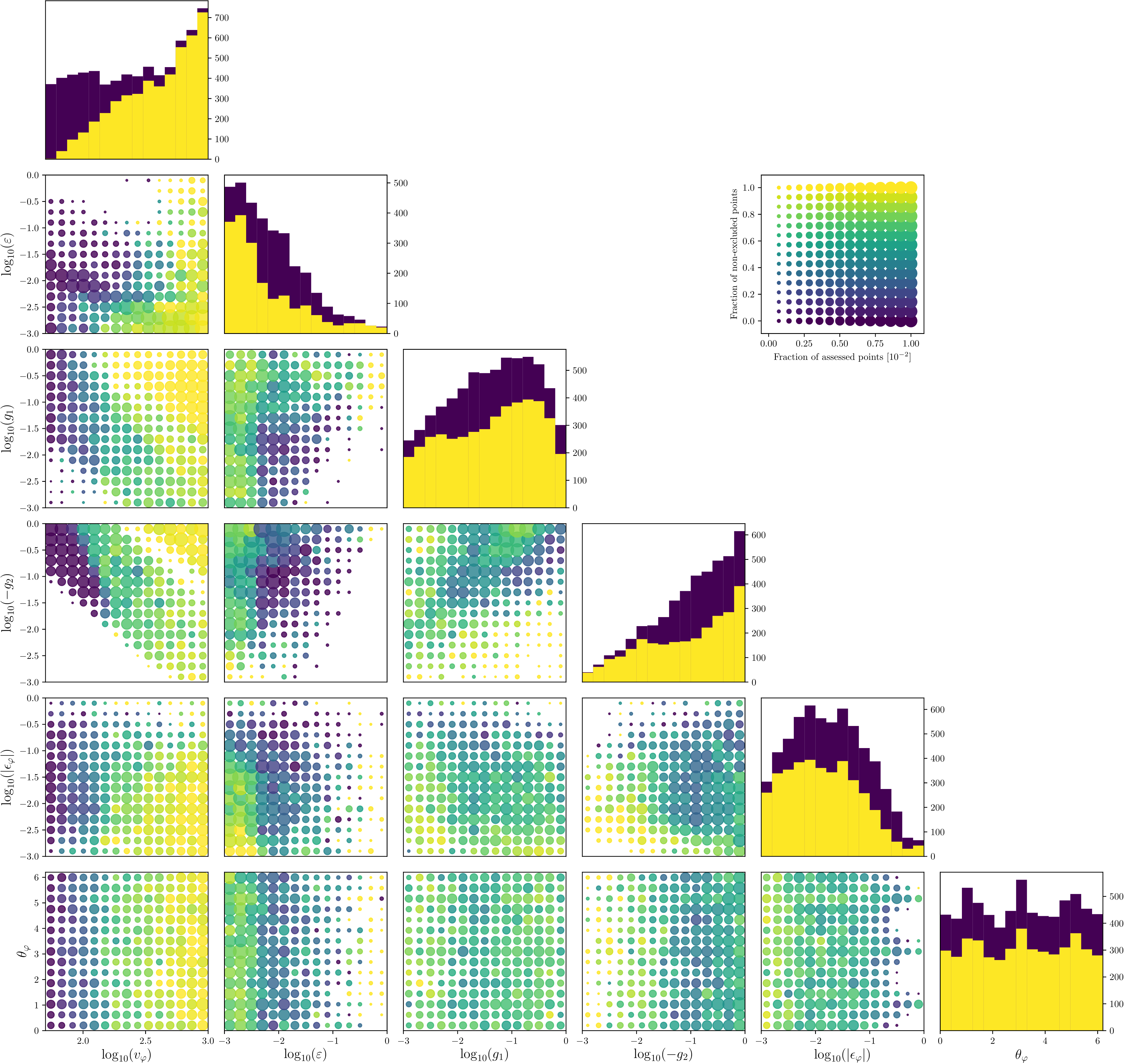}
    \caption{Results for \atlas only excluding points for which $v_{\varphi}<10^{1.8}$ GeV.  As before  the histograms show the 1-dimensional projections of the number of excluded (purple) and not excluded (yellow) points. The scatter plots are a representation of 2-dimensional projections. }
    \label{fig:ATLASexcluded}
\end{figure}
\begin{figure}[ht]
    \centering
    \includegraphics[width=\textwidth]{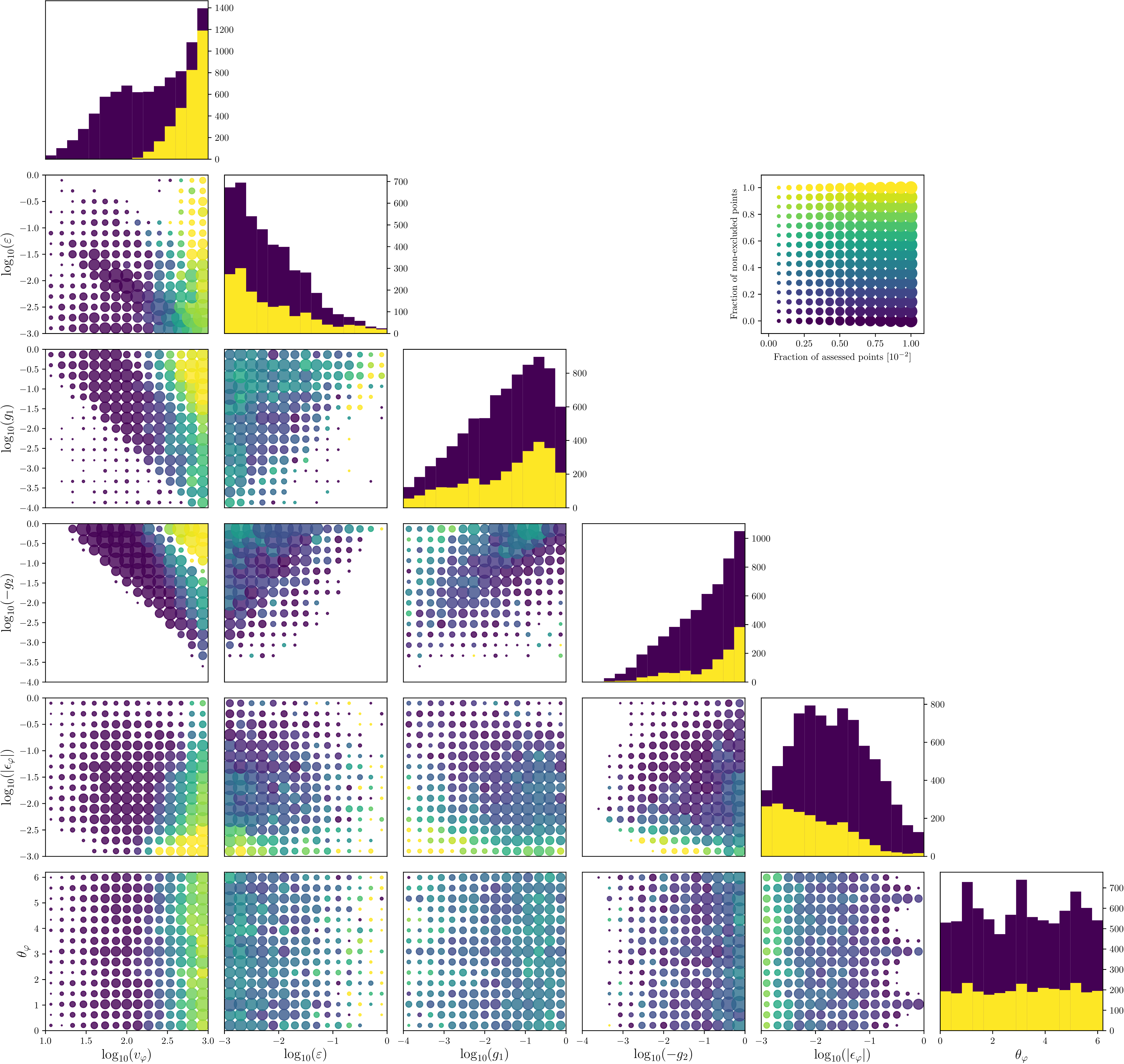}
    \caption{Results for \meg only. The histograms show the 1-dimensional projections of the number of excluded (purple) and not excluded (yellow) points. The scatter plots are a representation of 2-dimensional projections. The size of the circle indicates the fraction of the $N_\text{tot}$ points analysed in a single bin while the colour shows what fraction of those points can be considered excluded.}
    \label{fig:resultsMEG}
\end{figure}
\begin{figure}[ht]
    \centering
    \includegraphics[width=\textwidth]{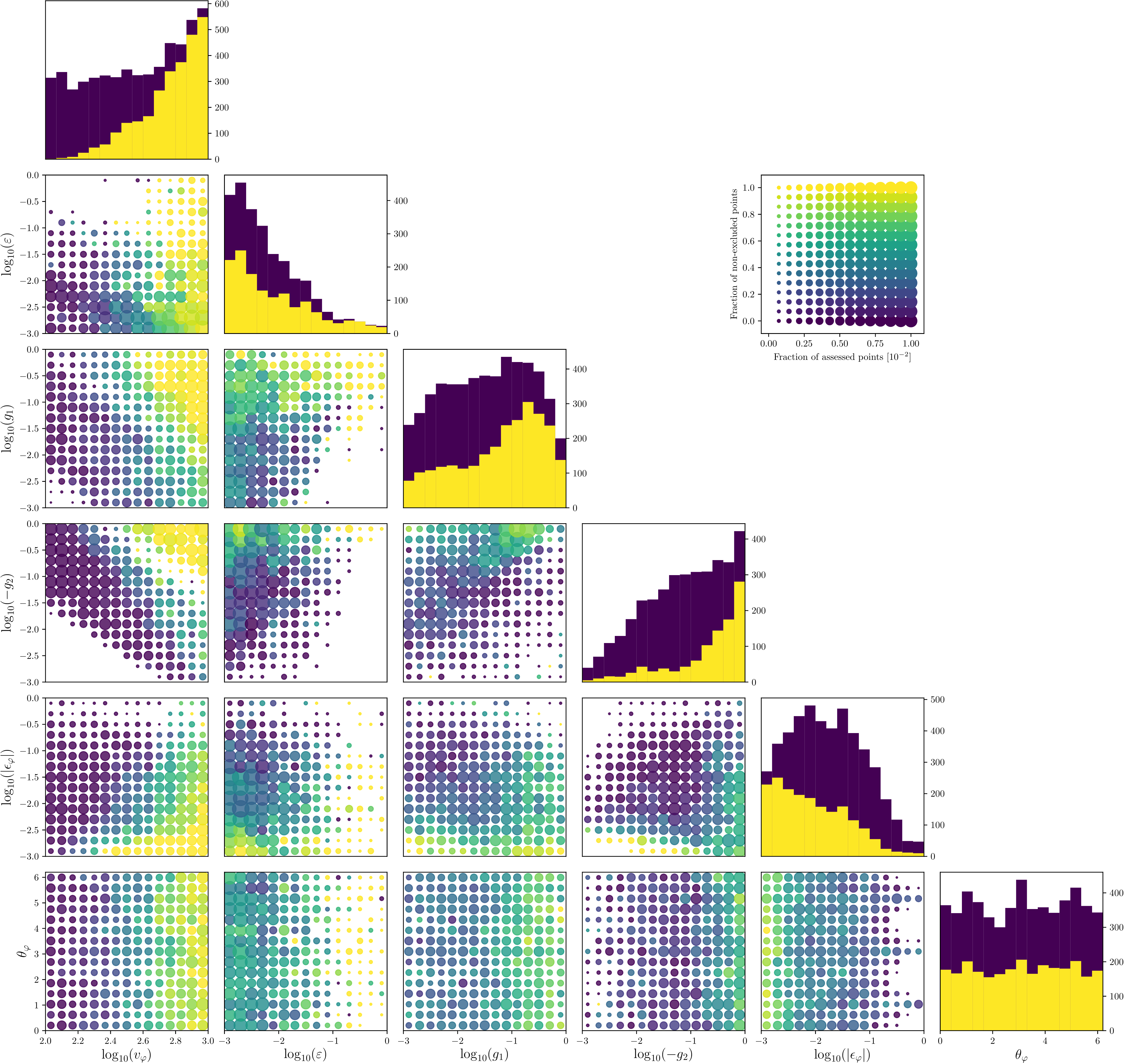}
    \caption{Results for \meg only excluding points for which $v_{\varphi}<10^{2.0}$ GeV. The histograms show the 1-dimensional projections of the number of excluded (purple) and not excluded (yellow) points. The scatter plots are a representation of 2-dimensional projections. The size of the circle indicates the fraction of the $N_\text{tot}$ points analysed in a single bin while the colour shows what fraction of those points can be considered excluded.}
    \label{fig:MEGexcluded}
\end{figure}
\begin{figure}[ht]
    \centering
    \includegraphics[width=\textwidth]{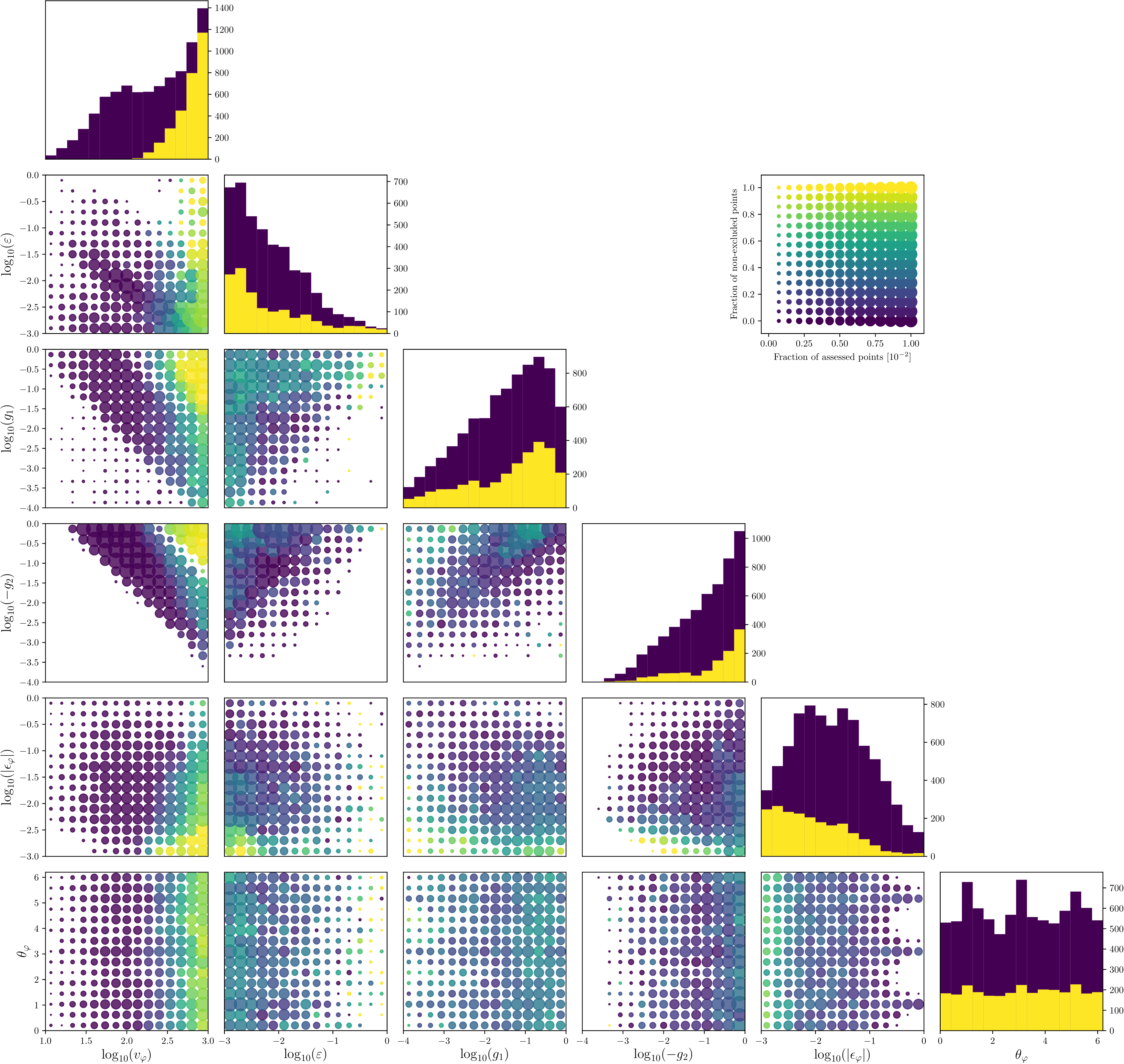}
    \caption{Results for \meg and \atlas combined. The histograms show the 1-dimensional projections of the number of excluded (purple) and not excluded (yellow) points. The scatter plots are a representation of 2-dimensional projections. The size of the circle indicates the fraction of the $N_\text{tot}$ points analysed in a single bin while the colour shows what fraction of those points can be considered excluded.}
    \label{fig:resultsBOTH}
\end{figure}
\begin{figure}[ht]
    \centering
    \includegraphics[width=\textwidth]{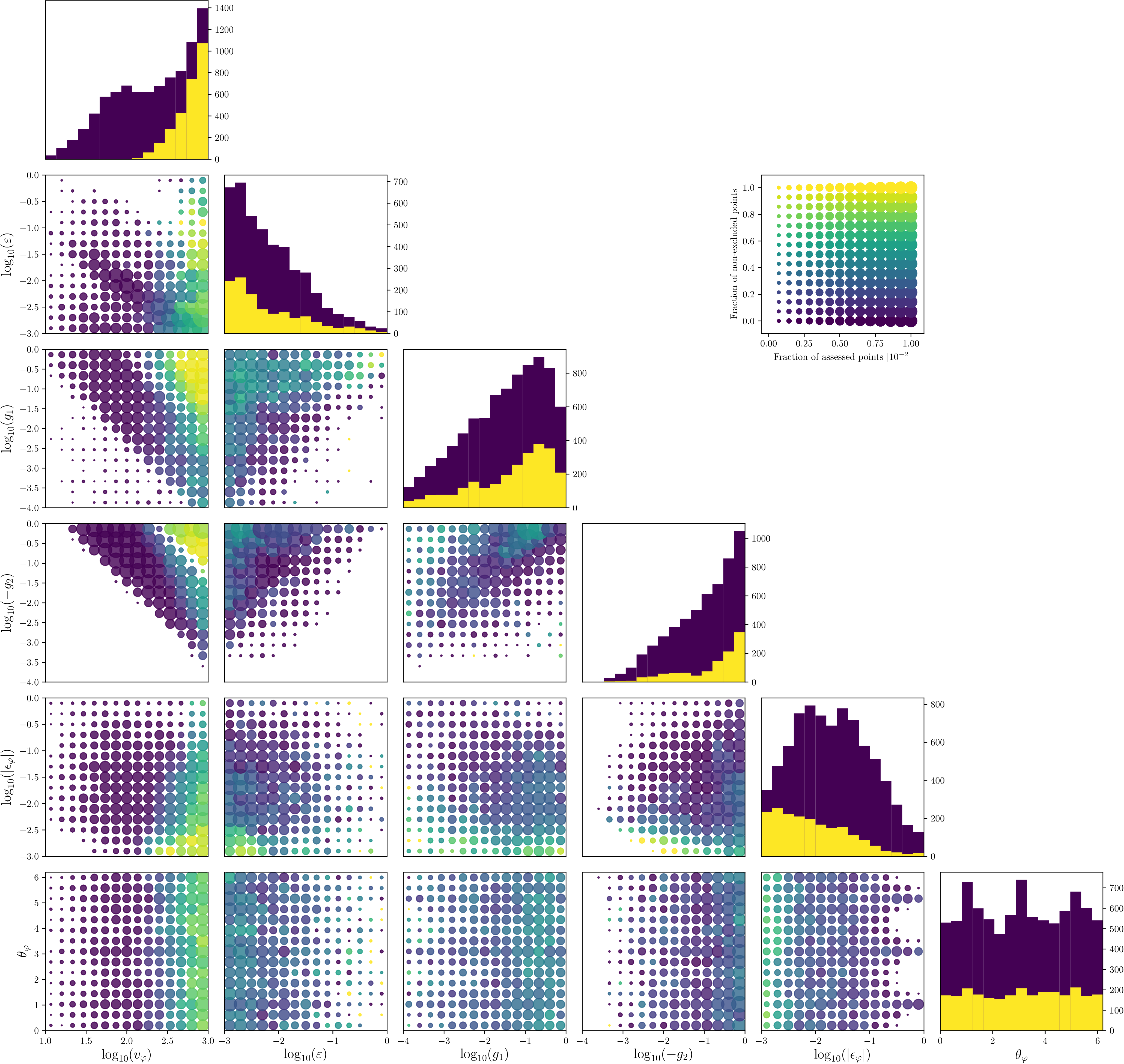}
    \caption{Results using all data constraints. The histograms show the 1-dimensional projections of the number of excluded (purple) and not excluded (yellow) points. The scatter plots are a representation of 2-dimensional projections. The size of the circle indicates the fraction of the $N_\text{tot}$ points analysed in a single bin while the colour shows what fraction of those points can be considered excluded.}
    \label{fig:resultsALL}
\end{figure}

%%%%%%%%%%%%%%%%%%%%%%%%%%%%%

The main results of this study are summarised in   \tabref{tab:exclusionfraction} and \tabref{tab:crossexclusion}. 
As can be seen from  \tabref{tab:exclusionfraction}, \meg excludes   $65.6\%$ of the total model parameter space, while the 8 TeV collider analysis excludes $40.0\%$. 
The constraints from Higgs measurements provide a total exclusionary power of $7.7\%$ while the \gmt experiment excludes the smallest 
volume of the parameter space. 

In addition to schematically quantifying the  exclusionary power of each measurement, we have demonstrated the
\emph{complementary} between the 
  collider analysis and that of \meg. As detailed in \tabref{tab:crossexclusion},
of the 3045 ($34.8\%$) points which cannot be excluded by \meg, we found  378 of those points ($12.4\%$) can be conservatively excluded by a combination of Higgs
width, Higgs-scalar mixing and the  ATLAS analysis. 
Moreover, of the 5317 ($60.0\%$) points which are not excluded by the collider analysis, 2386 ($44.87\%$) of those may be excluded by \meg.  
Interestingly, although
the  exclusionary power of the Higgs width and Higgs-scalar mixing is not considerable, when combined with the collider analysis, a sizeable portion of the 
parameter space which cannot be excluded by \meg becomes excluded. This is because the collider data is sensitive to the portal coupling of the flavons
with the Higgs. 

The exclusionary power of each experiment on the six-dimensional model parameter is presented in Figs.~\ref{fig:resultsATLAS}-\ref{fig:resultsALL}. 
These figures comprise of two types of plots: 
 two-dimensional projections of the six-dimensional model parameter space and histograms for a single model 
parameter. 
In the histograms the yellow regions represents the model parameter subspaces which
cannot be excluded by the relevant experiment while the dark blue denotes the regions which are excluded at $95\%$ C.L. 
In the two-dimensional projection plots the colours have the same meaning (yellow to dark blue represents lesser to greater  exclusionary power in those 
parameter space) but there is also additional information encoded in the size of the coloured dots: the larger the dot size the higher the density of sampled points in 
that region\footnote{Some regions are more sparsely populated from as the six conditions on the sampling space (\secref{sec:sampling})
can exclude more or less points in a given region of parameter space.}.

 \figref{fig:resultsATLAS} shows the results using \atlas data exclusively. We observe that for higher values of $v_{\varphi}$ (the VEV of the flavons) the 
 exclusionary power decreases. This is to be expected as the flavon masses increase with $v_{\varphi}$ and therefore their production cross section decreases. However, for the majority of the model 
parameter space, $v_{\varphi}\lesssim10^{2.0}$ GeV is excluded at $95\%$ C.L. In the case of the cross-coupling between the flavons and the Higgs, $\epsilon$, 
the smaller values  ($\epsilon<10^{-2}$) become increasingly more difficult to exclude. This is simply due to the fact that the production cross section of
the flavons decreases for smaller values of $\epsilon$. Moreover, we note that there is periodic behaviour in the polar coordinate, $\theta_{\varphi}$. 
%We observe that the exclusionary power in regards to $g_{1}$ and $g_{2}$ is not particularly strong.

We note that the excluded parameter-space constitutes an irregular body that does not align with 
the parameter axes. We therefore observe non-trivial correlations between the parameters
when testing for compatibility with experimental observation.  In an attempt to disentangle 
 which parameters are dominantly probed by  the ATLAS experiment we have 
plotted the regions of exclusion (for ATLAS) in which the points  $v<10^{1.8}$ GeV
    have been removed as shown in \figref{fig:ATLASexcluded}. From this figure, we observe that
    the shape of histograms for parameters $\epsilon$ and $\lvert
    \epsilon_{\varphi}\rvert$ changes relatively little after excluding the
    points for which $v_{\varphi}<10^{1.8}$ GeV. This implies  the ATLAS
    analysis has greatest sensitivity to these two parameters. In terms of the
    former parameter, $\epsilon$, this is the  cross coupling of the flavons to
    the  SM Higgs.  As the flavon production is directly mediated via this
    coupling, this explains the sensitivity of this ATLAS search to $\epsilon$.
    Moreover, as this ATLAS search looked for  final states which violated
    charged lepton flavour this analysis has sensitivity to $\lvert
    \epsilon_{\varphi}\rvert$. 
    
The exclusion from  \meg alone is shown in \figref{fig:resultsMEG}. It can be observed that the exclusionary power on $v_{\varphi}$,
which is the parameter that encapsulates the flavour breaking scale, is greater than that of \atlas alone and that $v_{\varphi}\lesssim10^{2.0}$ GeV is excluded at $95\%$ C.L.
Moreover, the cross-coupling is particularly constrained to a corner of the parameter space, $\epsilon<10^{-2.5}$.
We note that the constraints on $g_{1}$ and $g_{2}$ are much more aggressive compared to the constraint from \atlas.

%As there are non-trivial correlations between the parameters of the
%    model, in order to disentangle the impact of $v_{\varphi}$, 
%       we have shown
%    the same plot as \figref{fig:resultsMEG} but  with   the points  $v_{\varphi}<10^{2.0}$ GeV
%    excluded in \figref{fig:MEGexcluded}.  In general, the shape of the histograms of the other five
%    variables changes slightly.  However, the histograms which change least in
%    shape indicate which parameters are relatively  independent of
%    $v_{\varphi}$. Qualitatively, we find that the structure of the histogram
%    of $g_{2}$ and $\lvert\epsilon_{\varphi}\rvert$ changes the least which
%    implies MEG has sensitivity to these parameters in addition to that of
%    $v_{\varphi}$.  We expect MEG to have sensitivity to $g_{2}$ as the mass of
%    $\varphi_{2}$ is approximately proportional to  this parameter.  We observe this
%    to be the case and note that for larger values of $g_{2}$ (heavier masses
%    of $\varphi_{2}$) the exclusionary power of MEG becomes weaker as
%    anticipated.  Moreover, as the very CLFV nature of this model is
%parametrised by $\lvert\epsilon_{\varphi}\rvert$, it is unsurprising that MEG
%has high sensitivity to this parameter. 

As there are non-trivial correlations between the parameters of the
    model, in order to disentangle the impact of $v_{\varphi}$, 
       we show
    the same plot as \figref{fig:resultsMEG} but  with   the points  $v_{\varphi}<10^{2.0}$ GeV
    excluded in \figref{fig:MEGexcluded}.  In general, the shape of the histograms of the other five
    variables changes slightly.  However, the histograms which change least in
    shape indicate which parameters are relatively  independent of
    $v_{\varphi}$. Qualitatively, we find that the structure of the histogram
    of $\lvert\epsilon_{\varphi}\rvert$ changes the least which
    implies MEG has sensitivity to this parameters in addition to that of
    $v_{\varphi}$.  We note that the shape of the other histograms for parameters $g_{1}$, $g_{2}$, $\theta_{\varphi}$ 
    and $\epsilon$ all change significantly. This implies the exclusionary power of MEG in those parameters
    is correlated with $v_{\varphi}$. It is unsurprising that MEG has sensitivity to $\lvert\epsilon_{\varphi}\rvert$
    as  the very CLFV nature of this model is
parametrised by this variable.

The plot in \figref{fig:resultsBOTH} shows the combined \meg and \atlas constraints. Finally, \figref{fig:resultsALL} 
shows the constraints from all included experimental data. 
We find that there is not a significant qualitative difference between the two plots (as the Higgs and \gmt constraints are very weak).  We observe that the flavour breaking scale, parametrised by $v_{\varphi}$,  must be greater than $\sim 10^{2.5}$ GeV. Moreover, the cross-coupling
between this flavour sector and the SM must be $\epsilon<10^{-2}$. The absolute value of the parameter which controls how much 
the residual $\mathbb{Z}_3$ symmetry of the charged lepton sector is broken, is particularly constrained
$\lvert \epsilon_{\varphi}\rvert<10^{-2.75}$
\footnote{We note that this
statement naturally depends on the specific point in the model parameter space.}. However, the polar coordinate of the $\mathbb{Z}_3$-breaking parameter
 is constrained to $\theta_{\varphi}<0.5$ radians. In summary, the majority of the chosen model parameter space can be excluded through the combination of the measurements from the \atlas analysis, \meg, \gmt experiments and Higgs measurement data. 

%%%%%%%%%%%%%%%%%%%%%%%%%%%%%
%%%%%%%%  CONCLUSION      %%%%%%%
%%%%%%%%%%%%%%%%%%%%%%%%%%%%%
\section{Summary}\label{sec:summary}
 Explaining the origin of the flavour structure, 
in both quark and lepton sectors, has been a recurring theme in many proposed extensions of the SM. 
One such approach to explain the pattern of leptonic mixing is the application of discrete, non-Abelian 
flavour symmetries. This flavour symmetry must be broken in the low-energy effective theory
but its residual symmetries survive and play an important role in predicting the structure of the leptonic mixing matrix. 

In this work, we did not  presuppose the flavour breaking scale was high (close to the GUT scale) and therefore it is an interesting endeavour to constrain this flavour model's parameter space using a synergy of experimental data. 
In order to exclude regions of  the model parameter space, we applied constraints from \gmt, \meg, Higgs-scalar mixing and  Higgs width measurements. 
In addition, we recasted an 8 TeV \atlas analysis which searched for events with high-multiplicities of leptonic final states. 
We believe we are the first to combine, in both a conservative and rigorous manner, such experimental data to constrain a leptonic flavour model. 
 
One of our primary aims was to be as generic as possible in constraining the parameter space of a well-motivated flavour model, such as $A_4$. 
Therefore, we chose to  investigate  a simplified description of an $A_4$ model. 
At leading order this model has  the general features of most $A_4$ models, where the residual symmetry $\mathbb{Z_3}$ ($\mathbb{Z_2}$) is preserved in the charged lepton (neutrino) sector and consequently  TBM mixing is predicted after $A_4$ symmetry breaking. However, at sub-leading order, the residual symmetries are slightly broken 
due to the shift in one of the flavon VEVs.  As a consequence, a pattern of mixing consistent with neutrino oscillation data is achieved. 
We mainly focus on the phenomenology of the flavon, $\varphi$ which couples to charged leptons as it has greater experimental visibility than its neutrino flavon counterpart, $\chi$.

We conducted an exploration of the six-dimensional model parameter space and found most of the constraints could be calculated analytically.
The collider reinterpretation of an 8 TeV \atlas measurement was a more involved process. We benefited greatly  from the \atlas
collaboration both preserving and validating their analysis; moreover, the analysis was publicly available via \rivet which we used as our analysis tool.
 We believe we are the first 
to utilise the Monte-Carlo event generator, \sherpa, for BSM purposes, which was particularly amenable as the fully-showered and hadronised Monte-Carlo events could 
directly be fed into the analysis tool.  
Although we focused on an economical  model, most basic features of leptonic flavour models have been included, e.g., interactions related to lepton mass generation, interactions leading to the breaking of flavour symmetry and residual symmetries, as well as the Higgs-portal interactions. An alternative model may increase the number of free parameters but preserve most of these features and this investigation remain relevant.

We found the most aggressive constraints derived from the CLFV limit set by the \meg experiment; approximately $\sim60\%$ of the 
parameter space could be excluded.  The exclusionary power exhibited by the \atlas analysis came second only to \meg; excluding $\sim40\%$.
The remaining experimental data had smaller but non-negligible exclusionary power. Interestingly, the exclusionary power of  \meg and and the collider experimental
data complement each other: the collider
analysis, combined with Higgs width and Higgs-scalar mixing constraints,  can exclude regions \meg simply cannot and vice versa. This is because the collider search is sensitive to the mixing of the flavons
with the SM Higgs doublet, while the constraints from \meg is not. 

We hope the collider experimentalists view this optimistically, that searches for high-multiplicity leptonic final states
exclude sizeable regions of the leptonic flavour  model parameter space and 
complement
limits set by experiments dedicated to searching for CLFV. Moreover, at higher centre of mass energies, the exclusionary power of collider searches for anomalous production of
leptonic final states may seriously compete  
with those of \meg. Work to precisely quantify this statement is of interest but relegated for future studies. 
In addition,  the construction and optimisation of an analysis for
 final states such as $\tau\tau\mu e$, which would be a useful step in constraining the flavour breaking scale, if indeed such a scale exists.

\acknowledgments
We would like to thank  Bogdan Dobrescu for helpful advice throughout and proofreading this manuscript. 
We are grateful to Beate Heinemann and Mike Hance for useful advice 
regarding the reinterpretation of  ATLAS analysis presented in \cite{Aad:2014hja}. It is a pleasure to thank Silvan Kuttimalai on helpful discussions regarding Sherpa. We would like to thank Zhen Liu,
Serguey Petcov   and Alexis Plascencia for helpful conversations about various aspects of this work.  L.H. is supported through NSF ACI-1450310 and PHY-1505463. This manuscript has been authored by Fermi Research Alliance, LLC under Contract No. DE-AC02-07CH11359 with the U.S. Department of Energy, Office of Science, Office of High Energy Physics. This material is based upon work supported by the U.S. Department of Energy, Office of Science, Office of Advanced Scientific Computing Research, Scientific Discovery through Advanced Computing (SciDAC) program.
Y.L.Z. acknowledges the STFC Consolidated Grant ST/L000296/1 and the European Union's Horizon 2020 Research and Innovation programme under Marie Sk\l{}odowska-Curie grant agreements Elusives ITN No.\ 674896 and InvisiblesPlus RISE No.\ 690575.

\newpage
\appendix

\section{Group Theory for $A_4$}\label{sec:A_4reps}
In this  paper, as in \cite{Pascoli:2016wlt}, we work in the Altarelli-Feruglio basis \cite{Altarelli:2005yx}, where $T$ is diagonal. $T$ and $S$ are respectively given by  
\begin{eqnarray}
T=\left(
\begin{array}{ccc}
 1 & 0 & 0 \\
 0 & \omega ^2 & 0 \\
 0 & 0 & \omega  \\
\end{array}
\right)\,, \qquad
S=\frac{1}{3} \left(
\begin{array}{ccc}
 -1 & 2 & 2 \\
 2 & -1 & 2 \\
 2 & 2 & -1 \\
\end{array}
\right) \,.
\label{eq:generators}
\end{eqnarray}
 The products of two 3-dimensional irreducible representations $a=(a_1,a_2,a_3)^T$ and $b=(b_1,b_2,b_3)^T$ can be expressed as
\begin{equation}
\begin{aligned}
(ab)_\mathbf{1} &= a_1b_1 + a_2b_3 + a_2b_3 \,\\
(ab)_\mathbf{1'}&= a_3b_3 + a_1b_2 + a_2b_1 \,\\
(ab)_\mathbf{1''} &= a_2b_2 + a_1b_3 + a_3b_1 \,\\
(ab)_{\mathbf{3}_S} &= \frac{1}{2} \left(\begin{array}{c} 2a_1b_1-a_2b_3-a_3b_2\\ 2a_3b_3-a_1b_2-a_2b_1\\ 2a_2b_2-a_3b_1-a_1b_3\end{array} \right),
(ab)_{\mathbf{3}_A} = \frac{1}{2} \left(\begin{array}{c} a_2b_3-a_3b_2\\ a_1b_2-a_2b_1\\ a_3b_1-a_1b_3 \end{array} \right) .
\label{eq:CG2}
\end{aligned}
\end{equation}

\section{Higgs Width Constraint}
In this Appendix, we provide the coupling of the Higgs to the flavon mass eigenstates and detail how the SM Higgs width is calculated from such coupling. 
As introduced in \secref{sec:modelandint}, the gauge and mass eigenstates of the scalar sector in the $\mathbb{Z}_3$-breaking scenario, may be related by the unitary matrix 
\[
\begin{pmatrix}
\tilde{h}\\
\tilde{\varphi}_1\\
\sqrt{2}\text{Re}\left( \varphi_2\right)\\
\sqrt{2}\text{Im}\left( \varphi_2\right)
\end{pmatrix}=
\begin{pmatrix}
W_{00} & W_{01}& W_{02} & W_{03}\\
W_{10} & W_{11}& W_{12} & W_{13}\\
W_{20} & W_{21}& W_{22} & W_{23}\\
W_{30} & W_{31}& W_{32} & W_{33}\\
\end{pmatrix}
\begin{pmatrix}
h\\
s_1\\
s_2\\
s_3
\end{pmatrix}.
\]
Using the above notation, the $\mathbb{Z}_3$-breaking triplet couplings  are given below
\begin{equation}
\begin{aligned}
  \tilde{g}_{hs^2_1} & = g_2 v_{\varphi}W_{10} W_{11}^2
					  -\frac{1}{2} g_2 v_{\varphi}W_{20} W_{21}^2
			 		 -\frac{1}{2} g_2 v_{\varphi}W_{30} W_{31}^2
					 -\frac{1}{2} g_2 \epsilon_{\varphi}v_{\varphi} W_{10} W_{21}^2\\
& 					 -g_2  \epsilon_{\varphi}v_{\varphi} W_{11} W_{20} W_{21}
 					 +3 g_2 \epsilon_{\varphi}v_{\varphi} W_{20} W_{21}^2
 					 -\frac{1}{2} g_2 \epsilon^*_{\varphi}v_{\varphi} W_{10} W_{31}^2
 					 -g_2 \epsilon^*_{\varphi}v_{\varphi} W_{11} W_{30} W_{31}\\
&				 	 +3 \lambda  v_{H}W_{00} W_{01}^2
				 	 +\frac{1}{2} v_{H}W_{00} W_{11}^2 \epsilon 
				 	 +v_{H}W_{00} W_{21} W_{31} \epsilon 
					  +v_{H} W_{01} W_{10} W_{11} \epsilon \\
  &					 +v_{H} W_{01} W_{20} W_{31} \epsilon 
 					  +v_{H} W_{01} W_{21} W_{30} \epsilon 
				  	 +v_{\varphi}W_{00} W_{01}  W_{11} \epsilon 
 					  +\frac{1}{2} v_{\varphi}W_{01}^2 W_{10} \epsilon \\
 &					  +\epsilon_{\varphi}v_{\varphi}W_{00} W_{01}  W_{31} \epsilon 
 					  +\frac{1}{2} \epsilon_{\varphi}v_{\varphi} W_{01}^2 W_{30} \epsilon
 				   +\epsilon^*_{\varphi}v_{\varphi}W_{00} W_{01} W_{21} \epsilon 
 				   +\frac{1}{2} \epsilon^*_{\varphi}v_{\varphi} W_{01}^2 W_{20} \epsilon,
				   	\end{aligned}
	\end{equation}
	\begin{equation}
	\begin{aligned}
 	  \tilde{g}_{hs^2_2} & = g_2 v_{\varphi}W_{10} W_{12}^2
	  -\frac{1}{2} g_2 v_{\varphi}W_{20} W_{22}^2
	  -\frac{1}{2} g_2 v_{\varphi}W_{30} W_{32}^2
	  -\frac{1}{2} g_2 \epsilon_{\varphi}v_{\varphi} W_{10} W_{22}^2\\
&	  -g_2   \epsilon_{\varphi}v_{\varphi} W_{12} W_{20} W_{22}
	  +3 g_2 \epsilon_{\varphi}v_{\varphi} W_{20} W_{22}^2
	  -\frac{1}{2} g_2   \epsilon^*_{\varphi}v_{\varphi} W_{10} W_{32}^2
	  -g_2 \epsilon^*_{\varphi}v_{\varphi} W_{12} W_{30} W_{32} \\
&	  +3 \lambda  v_{H} W_{00} W_{02}^2
	  +\frac{1}{2} v_{H}W_{00} W_{12}^2 \epsilon
	   +v_{H}W_{00} W_{22}W_{32} \epsilon
	    +v_{H} W_{02} W_{10} W_{12} \epsilon \\
&	    +v_{H} W_{02} W_{20}  W_{32} \epsilon 
	    +v_{H} W_{02} W_{22} W_{30} \epsilon 
	    +v_{\varphi}W_{00} W_{02}W_{12} \epsilon
	     +\frac{1}{2} v_{\varphi}W_{02}^2 W_{10} \epsilon \\
&	     +\epsilon_{\varphi}v_{\varphi}W_{00} W_{02}W_{32} \epsilon
	      +\frac{1}{2} \epsilon_{\varphi}v_{\varphi} W_{02}^2 W_{30} \epsilon 
	      +\epsilon^*_{\varphi}v_{\varphi}W_{00} W_{02}W_{22} \epsilon,
	      +\frac{1}{2} \epsilon^*_{\varphi}v_{\varphi} W_{02}^2 W_{20} \epsilon,
	         \end{aligned}
   \end{equation}
   \begin{equation}
   \begin{aligned}
    \tilde{g}_{hs^2_3} & =g_2 v_{\varphi}W_{10} W_{13}^2
     -\frac{1}{2} g_2 v_{\varphi}W_{20}W_{23}^2
    -\frac{1}{2}g_2 v_{\varphi}W_{30}W_{33}^2
    -\frac{1}{2} g_2 \epsilon_{\varphi}v_{\varphi} W_{10}W_{23}^2\\
&    -g_2\epsilon_{\varphi}v_{\varphi} W_{13} W_{20}W_{23}
    +3 g_2 \epsilon_{\varphi}v_{\varphi} W_{20}W_{23}^2
    -\frac{1}{2} g_2\epsilon^*_{\varphi}v_{\varphi} W_{10}W_{33}^2
    -g_2 \epsilon^*_{\varphi}v_{\varphi} W_{13} W_{30}W_{33} \\
&    +3 \lambda  v_{H}W_{00} W_{03}^2
    +\frac{1}{2} v_{H}W_{00} W_{13}^2 \epsilon 
    +v_{H}W_{00}W_{23}W_{33} \epsilon 
    +v_{H} W_{03} W_{10} W_{13} \epsilon \\
&     +v_{H} W_{03} W_{20} W_{33} \epsilon 
     +v_{H} W_{03}W_{23} W_{30} \epsilon
      +v_{\varphi}W_{00} W_{03} W_{13} \epsilon 
      +\frac{1}{2} v_{\varphi}W_{03}^2 W_{10} \epsilon \\
&      +\epsilon_{\varphi}v_{\varphi}W_{00} W_{03}W_{33} \epsilon 
      +\frac{1}{2} \epsilon_{\varphi}v_{\varphi} W_{03}^2 W_{30} \epsilon
       +\epsilon^*_{\varphi}v_{\varphi}W_{00} W_{03}W_{23} \epsilon
        +\frac{1}{2} \epsilon^*_{\varphi}v_{\varphi} W_{03}^2 W_{20} \epsilon,
   \end{aligned}
\end{equation}
  \begin{equation}
   \begin{aligned}
    \tilde{g}_{hs_1s_2} & = 
	2 g_{2} v_{\varphi} W_{10} W_{11} W_{12} - g_{2} v_{\varphi} W_{20} W_{21}
 	  W_{22}-g_{2} v_{\varphi} W_{30} W_{31} W_{32}-g_{2} \epsilon_{\varphi}v_{\varphi} W_{10} W_{21}
 	  W_{22}\\
	  & -g_{2} \epsilon_{\varphi}v_{\varphi} W_{11} W_{20} W_{22}-g_{2} \epsilon_{\varphi}v_{\varphi} W_{12} W_{20}
 	  W_{21}+6 g_{2} \epsilon_{\varphi}v_{\varphi} W_{20} W_{21} W_{22}-g_{2} \epsilon^*_{\varphi}v_{\varphi} W_{10} W_{31}W_{32}\\
	  &-g_{2} \epsilon^*_{\varphi}v_{\varphi} W_{11} W_{30} W_{32}-g_{2} \epsilon^*_{\varphi}v_{\varphi} W_{12} W_{30}
 	  W_{31}+6 \lambda  v_{H} W_{00} W_{01} W_{02}+v_{H} W_{00} W_{11} W_{12}  \epsilon \\
	  &+v_{H} W_{00} W_{21} W_{32} \epsilon +v_{H} W_{00} W_{22} W_{31}  \epsilon +v_{H} W_{01} W_{10} W_{12} \epsilon +v_{H} W_{01} W_{20} W_{32}
 	  \epsilon \\
	  &+v_{H} W_{01} W_{22} W_{30} \epsilon +v_{H} W_{02} W_{10} W_{11}
 	  \epsilon +v_{H} W_{02} W_{20} W_{31} \epsilon +v_{H} W_{02} W_{21} W_{30}
	   \epsilon \\
	   &+v_{\varphi} W_{00} W_{01} W_{12} \epsilon +v_{\varphi} W_{00} W_{02} W_{11}
 	  \epsilon +v_{\varphi} W_{01} W_{02} W_{10} \epsilon +\epsilon_{\varphi}v_{\varphi} W_{00} W_{01} W_{32}
	   \epsilon \\
	   &+\epsilon_{\varphi}v_{\varphi} W_{00} W_{02} W_{31} \epsilon +\epsilon_{\varphi}v_{\varphi} W_{01} W_{02} W_{30}
	   \epsilon +\epsilon^*_{\varphi}v_{\varphi} W_{00} W_{01} W_{22} \epsilon +\epsilon^*_{\varphi}v_{\varphi} W_{00} W_{02} W_{21}
	   \epsilon\\
	   & +\epsilon^*_{\varphi}v_{\varphi} W_{01} W_{02} W_{20} \epsilon,
         \end{aligned}
\end{equation}
  \begin{equation}
   \begin{aligned}
   \tilde{g}_{hs_1s_3} & =  
		2 g_{2}v_{\varphi} W_{10} W_{11} W_{13} - g_{2}v_{\varphi} W_{20} W_{21}
  		 W_{23} - g_{2}v_{\varphi} W_{30} W_{31} W_{33}-g_{2}\epsilon_{\varphi}v_{\varphi} W_{10} W_{21}
   W_{23}\\
   &-g_{2}\epsilon_{\varphi}v_{\varphi} W_{11} W_{20} W_{23}-g_{2}\epsilon_{\varphi}v_{\varphi} W_{13} W_{20}
   W_{21}+6 g_{2}\epsilon_{\varphi}v_{\varphi} W_{20} W_{21} W_{23}-g_{2}\epsilon^*_{\varphi}v_{\varphi} W_{10} W_{31}
   W_{33}\\
   &-g_{2}\epsilon^*_{\varphi}v_{\varphi} W_{11} W_{30} W_{33} -g_{2}\epsilon^*_{\varphi}v_{\varphi} W_{13} W_{30}
   W_{31} + 6 \lambda  v_{H} W_{00} W_{01} W_{03} + v_{H} W_{00} W_{11} W_{13}
   \epsilon \\
   &+v_{H} W_{00} W_{21} W_{33} \epsilon +v_{H} W_{00} W_{23} W_{31}
   \epsilon +v_{H} W_{01} W_{10} W_{13} \epsilon +v_{H} W_{01} W_{20} W_{33}
   \epsilon \\
   & +v_{H} W_{01} W_{23} W_{30} \epsilon +v_{H} W_{03} W_{10} W_{11}
   \epsilon +v_{H} W_{03} W_{20} W_{31} \epsilon +v_{H} W_{03} W_{21} W_{30}
   \epsilon \\
   & +v_{\varphi} W_{00} W_{01} W_{13} \epsilon +v_{\varphi} W_{00} W_{03} W_{11}
   \epsilon +v_{\varphi} W_{01} W_{03} W_{10} \epsilon +\epsilon_{\varphi}v_{\varphi} W_{00} W_{01} W_{33}
   \epsilon \\
   &+\epsilon_{\varphi}v_{\varphi} W_{00} W_{03} W_{31} \epsilon +\epsilon_{\varphi}v_{\varphi} W_{01} W_{03} W_{30}
   \epsilon +\epsilon^{*}_{\varphi}v_{\varphi} W_{00} W_{01} W_{23} \epsilon +\epsilon^{*}_{\varphi}v_{\varphi} W_{00} W_{03} W_{21}
   \epsilon \\
   &+\epsilon^{*}_{\varphi}v_{\varphi} W_{01} W_{03} W_{20} \epsilon,
            \end{aligned}
\end{equation}
  \begin{equation}
   \begin{aligned}
      \tilde{g}_{hs_2s_3} & =  
	2 g_{2} v_{\varphi} W_{10} W_{12} W_{13} - g_{2} v_{\varphi} W_{20} W_{22}
   W_{23} - g_{2} v_{\varphi}W_{30} W_{32} W_{33} - g_{2} \epsilon_{\varphi}v_{\varphi} W_{10} W_{22}
   W_{23}\\
   &-g_{2} \epsilon_{\varphi}v_{\varphi} W_{12} W_{20} W_{23} - g_{2} \epsilon_{\varphi}v_{\varphi} W_{13} W_{20}
   W_{22} + 6 g_{2} \epsilon_{\varphi}v_{\varphi} W_{20} W_{22} W_{23} - g_{2} \epsilon^*_{\varphi}v_{\varphi} W_{10} W_{32}
   W_{33}\\
   &-g_{2} \epsilon^*_{\varphi}v_{\varphi} W_{12}W_{30} W_{33}-g_{2} \epsilon^*_{\varphi}v_{\varphi} W_{13}W_{30}
   W_{32}+6 \lambda  v_{H} W_{00} W_{02}W_{03}+v_{H} W_{00} W_{12} W_{13}
   \epsilon \\
   &+v_{H} W_{00} W_{22} W_{33} \epsilon +v_{H} W_{00} W_{23} W_{32}
   \epsilon +v_{H} W_{02} W_{10} W_{13} \epsilon +v_{H} W_{02} W_{20} W_{33}
   \epsilon \\
   & +v_{H} W_{02} W_{23}W_{30} \epsilon +v_{H}W_{03} W_{10} W_{12}
   \epsilon +v_{H}W_{03} W_{20} W_{32} \epsilon +v_{H}W_{03} W_{22}W_{30}
   \epsilon \\
   &+v_{\varphi} W_{00} W_{02} W_{13} \epsilon +v_{\varphi} W_{00}W_{03} W_{12}
   \epsilon +v_{\varphi} W_{02}W_{03} W_{10} \epsilon +\epsilon_{\varphi}v_{\varphi} W_{00} W_{02} W_{33}
   \epsilon \\
   &+\epsilon_{\varphi}v_{\varphi} W_{00}W_{03} W_{32} \epsilon +\epsilon_{\varphi}v_{\varphi} W_{02}W_{03}W_{30}
   \epsilon +\epsilon^*_{\varphi}v_{\varphi} W_{00} W_{02} W_{23} \epsilon +\epsilon^*_{\varphi}v_{\varphi} W_{00}W_{03} W_{22}
   \epsilon \\
   &+\epsilon^*_{\varphi}v_{\varphi} W_{02}W_{03} W_{20} \epsilon.
      \end{aligned}
\end{equation}
In comparison with the  $\mathbb{Z}_3$ preserving scenario, this case has a 
 larger number of possible mass combinations which may alter the Higgs total width. This is because the 
 Higgs can decay to three distinct flavons with possible mixtures of flavons in the final state. In the scenario, a single scalar
 contributes to the Higgs width,
 $m_{s_{i}}< m_{H}/2$ with $ i=1,2,3 $ and this implies
 \[
\Gamma\left(h \rightarrow s_i s_i \right) \leq 18 \,\text{MeV}.
\]
In the possibility of two scalars are lighter than half the mass of the Higgs, $m_{i}< m_{H}/2$ with $ i=1,2,3 $
then the 

\begin{itemize}
\item $m_{i}< m_{H}/2$ with $ i=1,2,3 $
\[
\Gamma\left(h \rightarrow s_i s_i \right) \leq 18 \,\text{MeV}
\]
\item  $m_{i}, m_{j} < m_{H}/2$ with $ i,j=1,2,3$
\[
\Gamma\left(h\rightarrow s_i s_i \right) + \Gamma\left(h\rightarrow s_j s_j \right) +  \Gamma\left(h\rightarrow s_i s_j \right) \leq 18 \,\text{MeV}
\]
\item $m_{i}, m_{j}, m_{k} < m_{H}/2$ with $ i,j,k =1,2,3$
\[
\Gamma\left(h\rightarrow s_i s_i \right) + \Gamma\left(h\rightarrow s_j s_j \right) + \Gamma\left(h\rightarrow s_k s_k \right) +
\Gamma\left(h\rightarrow s_i s_j \right) + \Gamma\left(h\rightarrow s_i s_k\right)  + \Gamma\left(h\rightarrow s_j s_k\right)\leq 18 \,\text{MeV}
\]
\end{itemize}
where 
\[
\Gamma\left(h\rightarrow s_i s_i \right)=\frac{\tilde{g}_{hs^2_i}}{8\pi m_h}\left(1-\frac{4m^2_{s_i}}{m^2_h}   \right)^{\frac{1}{2}}\quad \text{and}\quad
\Gamma\left(h\rightarrow s_i s_j \right) = \frac{\lvert p^*\rvert }{2\pi m^2_{h}}g^2_{hs_is_j}
\]
with
\[
\lvert p^*\rvert = \frac{1}{2m_H}\sqrt{\left[m^2_H - \left(m_{i}+m_j\right)^2\right]\left[m^2_H - \left(m_{i}-m_j\right)^2\right]}
\]

\bibliographystyle{JHEP}
\bibliography{ref}{}
\end{document}